\pdfoutput=1
\documentclass[a4paper,12pt]{article}
\usepackage{amsfonts}
\usepackage{mathrsfs}
\usepackage{amsmath}
\usepackage{amssymb}
\usepackage{framed}
\usepackage[medium]{titlesec}
\usepackage{bm}
\usepackage{cite}                                                   
\usepackage[normalem]{ulem}
\usepackage{extarrows}
\usepackage{slashed}
\usepackage{isodateo}
\usepackage{graphicx}
\usepackage{xcolor}
\usepackage[bookmarksnumbered=true,bookmarksopen=true]{hyperref}
 \hypersetup{colorlinks,%
             linkcolor=[rgb]{0,0.3,0.6}, %
             citecolor=[rgb]{0,0.3,0.6}, %
             urlcolor=[rgb]{0,0.3,0.6}}
\usepackage[hmargin=.7in,vmargin=1.1in]{geometry}
\usepackage{indentfirst}
\usepackage{booktabs}

\newcommand{\FR}[2]{\displaystyle\frac{\,{#1}\,}{#2}}
\newcommand{\fr}[2]{\mbox{$\frac{\,{#1}\,}{#2}$}}
\newcommand{\n}{\nonumber}

\graphicspath{{fig/}}

\def\bge{\begin{equation}}
\def\ede{\end{equation}}
\def\bga{\begin{aligned}}
\def\eda{\end{aligned}}
\def\bgb{\begin{bmatrix}}
\def\edb{\end{bmatrix}}
\def\bgp{\begin{pmatrix}}
\def\edp{\end{pmatrix}}
\def\bgm{\begin{matrix}}
\def\edm{\end{matrix}}
\def\bgs{\begin{subequations}}
\def\eds{\end{subequations}}
\newcommand{\order}[1]{\mathcal{O}({#1})}
\def\di{{\mathrm{d}}}

\def\mb{\mathbf}

\def\pd{\partial}

\def\la{\langle}\def\ra{\rangle}

\def\to{\rightarrow}

\def\ii{\mathrm{i}}

\def\al{\alpha}
\def\be{\beta}
\def\ga{\gamma}
\def\de{\delta}
\def\ep{\epsilon}
\def\ka{\kappa}
\def\lam{\lambda}

\def\si{\sigma}

\def\aa{\mathsf{a}}
\def\bb{\mathsf{b}}
\def\cc{\mathsf{c}}
\def\dd{\mathsf{d}}

\def\Re{\mathrm{Re}\,}

\usepackage{mdframed}

\newmdenv[skipabove=0pt,%
          skipbelow=5pt,%
          leftmargin=0pt,%
          rightmargin=0pt,%
          innertopmargin=-5pt,%
          innerbottommargin=7pt,%
          innerleftmargin=2pt,%
          innerrightmargin=2pt,%
          splittopskip=0pt,%
          splitbottomskip=0pt,%
          linewidth=0pt,%
          nobreak=true]%
          {keyeqn2}

\newmdenv[backgroundcolor=gray!15,%
          skipabove=0pt,%
          skipbelow=5pt,%
          leftmargin=0pt,%
          rightmargin=0pt,%
          innertopmargin=-5pt,%
          innerbottommargin=7pt,%
          innerleftmargin=2pt,%
          innerrightmargin=2pt,%
          splittopskip=0pt,%
          splitbottomskip=0pt,%
          linewidth=0pt,%
          nobreak=true]%
          {keyeqn}

\usepackage{titlesec}          
\titleformat{\section}
{\normalfont\fontsize{15}{20}\bfseries}{\thesection}{1em}{}

\newcommand{\wt}[1]{\mkern 2mu \widetilde{\mkern -2mu #1 \mkern -2mu}\mkern 2mu}
\newcommand{\wh}[1]{\mkern 2mu \widehat{\mkern-2mu#1\mkern-2mu}\mkern 2mu}

\newcommand{\fnemail}[1]{\footnote{Email: \href{mailto:#1}{\nolinkurl{#1}}}}

\begin{document}


\title{\Large\textbf{Closed-Form Formulae for Inflation Correlators}\\[2mm]}

\author{Zhehan Qin$^{1\,}$\fnemail{qzh21@mails.tsinghua.edu.cn}~~~~~ and ~~~~~Zhong-Zhi Xianyu$^{1,2\,}$\fnemail{zxianyu@tsinghua.edu.cn}\\[5mm]
\normalsize{${}^{1}\,$\emph{Department of Physics, Tsinghua University, Beijing 100084, China}}\\
\normalsize{${}^{2}\,$\emph{Collaborative Innovation Center of Quantum Matter, Beijing 100084, China}}}

\date{}
\maketitle

\vspace{20mm}

\begin{abstract}
\vspace{10mm}

We derive exact and closed-form expressions for a large class of two-point and three-point inflation correlators with the tree-level exchange of a single massive particle. The intermediate massive particle is allowed to have arbitrary mass, spin, chemical potential, and arbitrary nonderivative or derivative couplings to external inflaton modes. We also allow the coupling coefficients to have arbitrary complex power dependences on the conformal time. Our results feature closed-form expressions involving only familiar special functions and without any infinite sums. This is achieved by an improved bootstrap method with a suitable change of variables. Our results cover a wide range of cosmological collider models and can be directly used for future phenomenological studies. Our results can also be used as basic building blocks for constructing more complicated inflation correlators.

\end{abstract}

\newpage
\tableofcontents

\newpage
\section{Introduction}

Recent years have witnessed new progress in the study of particle physics during the cosmic inflation. It was realized that current and upcoming cosmological observations have great potential in probing the structures of primordial scalar and tensor perturbations, and thereby could provide us a wealth of information about the high-energy physics at the inflation scale. Many studies in recent years have explored the possibility of probing high-scale new physics with future data, a program sometimes called the ``cosmological collider (CC) physics''\cite{Chen:2009we,Chen:2009zp,Baumann:2011nk,Chen:2012ge,Pi:2012gf,Noumi:2012vr,Gong:2013sma,Arkani-Hamed:2015bza,Chen:2015lza,Chen:2016nrs,Chen:2016uwp,Chen:2016hrz,Lee:2016vti,An:2017hlx,An:2017rwo,Iyer:2017qzw,Kumar:2017ecc,Chen:2017ryl,Tong:2018tqf,Chen:2018sce,Chen:2018xck,Chen:2018cgg,Chua:2018dqh,Wu:2018lmx,Saito:2018omt,Li:2019ves,Lu:2019tjj,Liu:2019fag,Hook:2019zxa,Hook:2019vcn,Kumar:2018jxz,Kumar:2019ebj,Alexander:2019vtb,Wang:2019gbi,Wang:2019gok,Wang:2020uic,Li:2020xwr,Wang:2020ioa,Fan:2020xgh,Aoki:2020zbj,Bodas:2020yho,Maru:2021ezc,Lu:2021gso,Sou:2021juh,Lu:2021wxu,Pinol:2021aun,Cui:2021iie,Tong:2022cdz,Reece:2022soh,Qin:2022lva,Chen:2022vzh,Cabass:2022rhr,Cabass:2022oap,Niu:2022quw,Niu:2022fki}.

The main observables of CC physics are the $n$-point $(n\geq 2)$ correlation functions of the primordial scalar or tensor fluctuations. We collectively call them \emph{inflation correlators}. They can be viewed as correlation functions of quantum fields living in the bulk of the inflationary spacetime, which is approximately de Sitter, but with their external legs pinned to the future boundary of the spacetime, namely, the end of inflation.  As key quantities connecting QFT predictions with the cosmological observations, the inflation correlators play crucial roles in CC physics, much like the Minkowskian scattering amplitudes to the collider physics. It is thus of central importance to have a good theoretical understanding of inflation correlators. Although our current knowledge about inflation correlators is still very preliminary compared to the much-developed scattering amplitudes in Minkowski spacetime, a lot of new results have emerged in recent years in both analytical and numerical frontiers \cite{Arkani-Hamed:2018kmz,Baumann:2019oyu,Baumann:2020dch,Pajer:2020wnj,Pajer:2020wxk,Cabass:2021fnw,Pimentel:2022fsc,Jazayeri:2022kjy,Qin:2022fbv,Xianyu:2022jwk,Wang:2022eop,Baumann:2022jpr,Sleight:2019mgd,Sleight:2019hfp,Sleight:2020obc,Sleight:2021iix,Sleight:2021plv,Wang:2021qez,Goodhew:2020hob,Melville:2021lst,Goodhew:2021oqg,DiPietro:2021sjt,Tong:2021wai,Bonifacio:2021azc,Hogervorst:2021uvp,Meltzer:2021zin,Heckelbacher:2022hbq,Gomez:2021qfd,Gomez:2021ujt,Baumann:2021fxj}. 

A particular aspect of studying inflation correlators is to find precise and explicit results, either numerical or analytical, to a range of key processes that were identified in the recent studies of CC phenomenologies. Until recently, many phenomenological studies have relied on very often unjustified approximations.  In general, the CC processes involve inflation correlators mediated by massive fields at both the tree and the loop levels, and the on-shell production of these massive particles during inflation can leave oscillatory signatures in the inflation correlators. While this is phenomenologically quite appealing, the computation of these massive processes is relatively difficult. In the diagrammatic approach in the Schwinger-Keldysh (SK) formalism \cite{Schwinger:1960qe,Keldysh:1964ud,Feynman:1963fq,Jordan:1986ug,Weinberg:2005vy,Chen:2017ryl}, the computation involves multi-layered and nested time integrals over products of special functions. 

Several methods have been developed in recent years that benefit analytical computations of inflation correlators. For example, as was pointed out in \cite{Arkani-Hamed:2015bza}, one can derive differential equations satisfied by inflation correlators, so that one can find explicit results by solving these differential equations with proper boundary conditions, instead of computing the SK integrals directly. This method has been further developed in subsequent studies and was called the cosmological bootstrap \cite{Arkani-Hamed:2018kmz,Baumann:2019oyu,Baumann:2020dch,Pajer:2020wnj,Pajer:2020wxk,Cabass:2021fnw,Pimentel:2022fsc,Jazayeri:2022kjy,Qin:2022fbv,Xianyu:2022jwk,Wang:2022eop,Baumann:2022jpr}. Borrowing this terminology and following our previous work \cite{Qin:2022fbv}, we shall call the differential equations satisfied by inflation correlators the \emph{bootstrap equations}. 

Another recently explored method is the Mellin transform \cite{Sleight:2020obc,Sleight:2021iix,Sleight:2021plv}, which exploits the dilatation symmetry of dS. The Mellin variable is essentially the weight of dilatation eigenmodes. By using Mellin variables in place of time variables, we can trivialize the SK time integrals. It was further shown in \cite{Qin:2022lva,Qin:2022fbv} that a more practical approach is to use Mellin variables only for the internal propagators, while the external legs still retain their time dependence. This approach, called partial Mellin-Barnes representation, is convenient for computing inflation correlators, since these correlators are defined to be equal-time correlators at the future boundary. It is thus better to leave the time variables in external modes untransformed. 

With either the bootstrap techniques or the Mellin transform, many explicit results have been found for tree-level correlators with single massive exchanges, including the dS covariant cases and boost-breaking cases. See, e.g., \cite{Arkani-Hamed:2018kmz,Baumann:2019oyu,Sleight:2019hfp,Pimentel:2022fsc,Jazayeri:2022kjy,Qin:2022fbv}. There are also new results at the 1-loop level obtained with partial Mellin-Barnes representation \cite{Qin:2022lva} or dS spectral decomposition \cite{Xianyu:2022jwk}. In all these examples, when the internal massive fields are heavy enough, the correlator contains an oscillatory/nonanalytic piece which we call the \emph{signal}, and a smooth/analytic piece which we call the \emph{background}. The signal part typically consists of imaginary powers of momentum ratios multiplied by a hypergeometric function, while the background part is typically written as a double Taylor series in momentum ratios. Most of the previous results were first obtained at the four-point level, and the three-point correlators were obtained by taking a soft limit, although taking this soft limit could be subtle in practice, as will be commented below.\footnote{There are also studies that work directly at the three-point level; See, e.g., \cite{Pimentel:2022fsc}.}  The results for the two-point functions are even rarer. 
 
\begin{figure}
\centering
  \parbox{0.38\textwidth}{\includegraphics[width=0.38\textwidth]{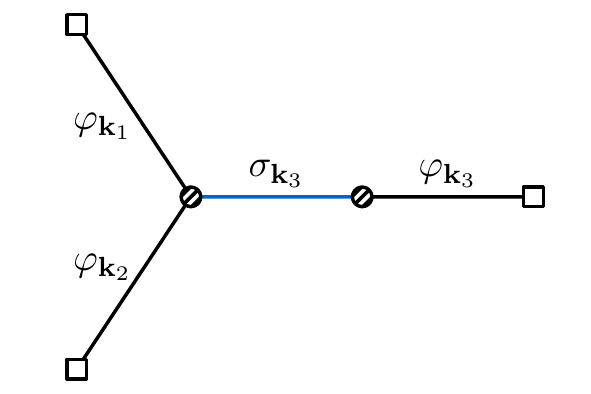}}~~~~
  \parbox{0.38\textwidth}{\includegraphics[width=0.38\textwidth]{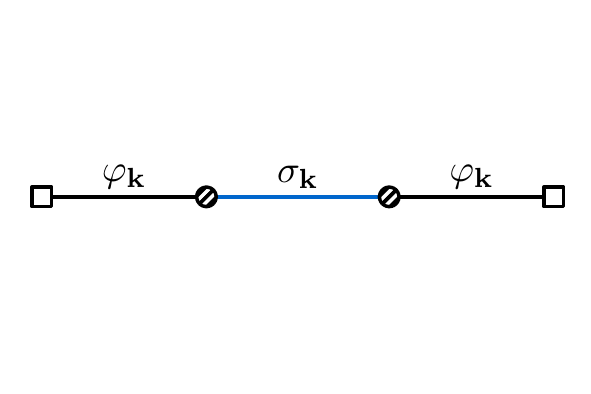}}
  \caption{Three-point correlator $\la\varphi_{\mb k_1}\varphi_{\mb k_2}\varphi_{\mb k_3}\ra'$ and two-point correlator $\la\varphi_{\mb k}\varphi_{-\mb k}\ra'$ mediated by a single massive field $\si$ at the tree level. Here $\varphi$ can be either the inflaton which is a massless scalar, or the graviton which is a massless spin-2 particle. Our diagrammatic notation follows \cite{Chen:2017ryl}. In particular, the shaded circles represent vertices of either SK branch $\aa=\pm$ which is summed over in the final result, and the white squares denote the boundary points at $\tau=0$. }
  \label{fd_3pt2pt}
\end{figure}

In this work, we present new analytical results for a wide range of three-point and two-point inflation correlators with a single massive exchange at the tree level, shown in Fig.\ \ref{fd_3pt2pt}. The external modes can be either massless scalar mode such as the inflaton fluctuation, or the closely related conformal scalar, or the massless tensor mode. The intermediate massive particle can have arbitrary mass, spin, and chemical potential within the physically allowed parameter space. We allow very general couplings between the external mode and the intermediate massive particle, including both the nonderivative and derivative couplings, and the coupling coefficient is allowed to have very general complex power dependence on the conformal time $\tau$. Thus our result covers models with time-dependent couplings, especially the oscillatory couplings that are present in models with oscillating background fields, e.g., \cite{Chen:2022vzh}. The couplings with arbitrary complex power dependence on the conformal time are also useful when we treat our correlators as subgraphs of more complicated correlators. 

Our results feature exact and closed-form formulae. In contrast to previously obtained results, our expressions for the three-point and two-point correlators do not contain any Taylor series. Instead, all the momentum dependences are fully captured by familiar special functions whose analytical properties are well understood. This makes our result particularly suitable for understanding the analytic properties of inflation correlators. Also, our compact results can also be used as handy building blocks when constructing more complicated inflation correlators, a topic we shall explore in a separate work \cite{qin_box}. 

The strategy we adopt in this paper is similar in spirit to many previous works on cosmological bootstraps. That is, we begin with a tree-level four-point correlator $\la\varphi_{\mb k_1}\varphi_{\mb k_2}\varphi_{\mb k_3}\varphi_{\mb k_4}\ra'$ with a single massive exchange in the $s$-channel, with momentum $\mb k_s\equiv \mb k_1+\mb k_2$. The result for the three-point and two-point correlators can be obtained by taking soft limits $\mb k_4\to 0$ and $\mb k_2,\mb k_4\to 0$, which we call the single folded limit and the double folded limit, respectively. (See Fig.\ \ref{fd_4pt}.) The regularity of folded limits of four-point functions follows from the Bunch-Davies initial condition for all the fields involved \cite{Arkani-Hamed:2018kmz}. So, there is no conceptual difficulty in taking the folded limits. In practice, however, the four-point functions usually contain terms singular in folded limits, and these singularities must cancel out in the full expression. The matter is further complicated by the fact that the double folded limit is usually at the boundary of convergent regions of the Taylor series for the background. All these complications make the folded limit less trivial than it seems. 

In this work, we circumvent these complications by making a proper change of variables. The advantage of adopting a new set of variables for the bootstrap equations has already been observed in \cite{Qin:2022fbv}. Here we shall make fuller use of it. The key idea is simple: after stripping off trivial external momentum factors, an $s$-channel four-point correlator depends on various momenta only through two independent momentum ratios. In previous works, the two independent momentum ratios are normally taken as $r_1\equiv k_{s}/(k_1+k_2)$ and $r_2\equiv k_s/(k_3+k_4)$, where $k_i\equiv |\mb k_i|$. Our new observation is that it is more advantageous to use $u_i\equiv 2r_i/(1+r_{i})$ $(i=1,2)$ instead of $r_{1,2}$. For physical parameters $0\leq r_{1,2}\leq 1$, the new variable $u_{i}$ is a monotonic function of $r_i$ and also takes its value from $[0,1]$. The key advantage of $u$ variables is that the inhomogeneous bootstrap equation can be solved by an ansatz of single-layer Taylor series in the single folded limit $u_2\to 1$, and it turns out that this Taylor series can be easily summed to give a closed-form special function, which in our case is always a generalized hypergeometric function ${}_3\text{F}_2$. With the closed-form expression for the three-point functions, the double folded limit $u_{1,2}\to 1$ is also easily taken, and the result is again a closed-form expression. 

Now we give an outline for the rest of this work. In Sec.\ \ref{sec_reduce}, we introduce the correlators to be computed in this work, and explain that all these correlators can be easily reduced to simple linear combinations of several \emph{seed integrals}, and thereby reduce the computation of the correlators to that of the seed integrals. Depending on whether there is a nonzero chemical potential for the intermediate massive particle, we need two types of seed integrals: For massive fields of arbitrary mass and spin but without chemical potential, the corresponding internal propagator can always be expressed in terms of Hankel functions. For such correlators, we define a \emph{Hankel seed integral}. On the other hand, for an intermediate field with a chemical potential, the transversely polarized internal propagator is expressed in terms of Whittaker W functions. For such cases, we define a \emph{Whittaker seed integral}. We then give various explicit examples, showing how to reduce the correlators to seed integrals. Then, we present all the details of solving the bootstrap equations for the Hankel seed integral in Sec.\ \ref{sec_hankel}, and for the Whittaker seed integral in Sec.\ \ref{sec_whittaker}. We conclude in Sec.\ \ref{sec_concl} with further discussions. Some useful formulae are summarized in App.\ \ref{app_formulae}, and a computation of the seed integrals in the squeezed limit is presented in App.\ \ref{app_squeezed}.

Readers uninterested in the technical details of bootstrapping seed integrals can skip most of Sec.\ \ref{sec_hankel} and Sec.\ \ref{sec_whittaker}. For readers who only want to use the results, the shortest route is to glance through Sec.\ \ref{sec_reduce} to get an idea of how to reduce a correlator to the seed integral, and then jump directly to the results. A quick summary of our results is as follows. The Hankel seed integral $\wt{\mathcal{I}}_{\aa\bb}^{p_1p_2}(u_1,u_2)$ is defined in (\ref{eq_SeedIntH}). The closed-form expression for $\wt{\mathcal{I}}_{\aa\bb}^{p_1p_2}(u_1,1)$ is given in (\ref{eq_IPPresult}) and (\ref{eq_IPMresult}), while the closed-form expression for $\wt{\mathcal{I}}_{\aa\bb}^{p_1p_2}(1,1)$ is given in (\ref{eq_H2ptResultPP}) and (\ref{eq_H2ptResultPM}). The Whittaker seed integral $\wt{\mathcal{I}}^{(h)p_1p_2}_{\aa\bb}(u_1,u_2)$ is defined in (\ref{eq_SeedIntW}). The closed-form expression for its single folded limit $\wt{\mathcal I}_{\aa\bb}^{(h)p_1p_2}(u,1)$ is given in (\ref{eq_IhPMresult}) and (\ref{eq_IhPPresult}), and the closed-form expression for the double folded limit $\wt{\mathcal I}_{\aa\bb}^{(h)p_1p_2}(1,1)$ is given in (\ref{eq_Ih2ptResultPM}) and (\ref{eq_Ih2ptResultPP}).

Explicit analytical results already exist in the literature for many special cases considered in this work, but almost always expressed in terms of Taylor series.\footnote{In the appendix of \cite{Arkani-Hamed:2018kmz} it was shown that such Taylor series in $r$ variables may be written in terms of the Kampé de Fériet function, which is a very general form of hypergeometric function of two variables. However, it seems that the numerical implementation of this function is not readily available in Mathematica.} For these known cases, our results are new in that they are expressed in compact and closed form in terms of familiar special functions, instead of Taylor series. Our results also include some completely new cases that were not treated in the literature, including most of two-point functions, the three-point function with (complex) time-dependent couplings, and also the chemical-potential-boosted three-point function with arbitrary couplings.

\paragraph{Notations and conventions.} We use mostly plus signature for the spacetime metric and we fix the background geometry to be the inflationary patch of dS. With the conformal time $\tau\in(-\infty,0)$ and comoving spatial coordinates $\mb x\in\mathbb R^3$, the metric reads $\di s^2 =a^2(\tau)(-\di\tau^2+\di\mb x^2)$ and $a(\tau)=-1/(H\tau)$, where $H$ is the inflation Hubble parameter. Throughout this work, we take $H=1$ for simplicity. We follow the diagrammatic notations and conventions in \cite{Chen:2017ryl}. Frequently used shorthand notations include $k_{ij}\equiv k_i+k_j$ $(i,j=1,2,3,4)$, $k_{123}\equiv k_1+k_2+k_3$, $p_{12}\equiv p_1+p_2$, $\bar p_{12}\equiv p_1-p_2$. Indices sans serif such as $\aa,\bb$ are often but not always labels for SK branches, but they always take values in $\pm 1$. In most places, we use the mass parameter $\wt\nu$ instead of the mass $m$ for the intermediate massive field. The two are related by $\wt\nu=\sqrt{m^2-9/4}$ for scalars and $\wt\nu=\sqrt{m^2-(s-1/2)^2}$ for fields of spin $s\neq 0$. In all mid-steps, we treat $\wt\nu$ as a positive real parameter. The results for complementary fields ($\wt\nu$ being purely imaginary) can be obtained in the final results by analytic continuation. For spinning fields, the chemical potential is denoted by $\wt\mu$. We use $\la\varphi_{\mb k_1}\cdots\varphi_{\mb k_n}\ra'$ to denote correlators of 3-momentum modes $\varphi_{\mb k}$, and the prime in $\la\cdots\ra'$ means that the momentum-conserving $\de$-function is removed. Other variables and parameters will be defined in the main text when they appear.

\section{Reducing Correlators to Seed Integrals}
\label{sec_reduce}

In this section, we first introduce the objects to be computed in this work, including the three-point and two-point correlators mediated by a single massive field at the tree level. Next, we introduce the two types of seed integrals, the Hankel seed integrals and the Whittaker seed integrals, which are built from the SK integrals in a standard diagrammatic calculation in the bulk. Then, we provide a number of examples showing how to reduce SK integrals to a simple linear combination of seed integrals, including both scalar and spinning exchanges with several types of couplings.

\subsection{Seed integrals: the definition and the result}

\paragraph{Ingredients: propagators and vertices.}
The inflation correlators considered in this work have two types of topologies shown in Fig.\ \ref{fd_3pt2pt}. That is, we consider correlators of the external field $\varphi$ mediated by a single massive field $\si$ at the three-point and two-point levels. The external field $\varphi$ can be a massless inflaton, a conformal scalar ($m^2=2$), or even a massless spin-2 graviton, since all these cases share very similar bulk-to-boundary propagators. Therefore, for the clarity of presentation, we will often take $\varphi$ as the massless inflaton field below, with a notable exception when we discuss the example of a massive spinning exchange with nonzero chemical potential. 

In the standard SK formalism, the bulk-to-boundary propagator of the inflaton field $\varphi$, represented by all the black external lines in Fig.\ \ref{fd_3pt2pt}, is given by \cite{Chen:2017ryl}:
\bge
\label{eq_Gbtob}
  G_\aa(k;\tau)=\FR{1}{2k^3}(1-\ii\aa k\tau)e^{\ii\aa k\tau}.
\ede
Here $k$ is the magnitude of the momentum flowing in the propagator. The conformal time $\tau$ and the SK index $\aa=\pm$ are both for the bulk endpoint. See \cite{Chen:2017ryl} for more detailed discussions. 

On the other hand, the blue internal lines in Fig.\ \ref{fd_3pt2pt} represent the bulk propagator of the massive field $\si$. In this work, the massive field is allowed to have either dS covariant or dS boost-breaking dispersion relation. The dS covariant case is technically simpler thanks to the presence of the full dS isometries, and is thus interesting on its own. The dS boost breaking case is more relevant to CC physics, since the rolling of the inflaton background necessarily breaks the dS boosts. More importantly, boost breaking processes typically lead to larger observable signals, and thus are more interesting for phenomenological model buildings \cite{Chen:2018xck,Hook:2019zxa,Liu:2019fag,Wang:2019gbi,Wang:2020ioa,Tong:2022cdz,Pimentel:2022fsc,Jazayeri:2022kjy,Qin:2022fbv,Niu:2022fki,Niu:2022quw}. 

Similar to the Minkowskian QFT, free massive particles with dS invariant dispersion relation are classified by their mass $m$ and spin $s$. The spin can take any nonnegative integer or half-integer value, although we shall only consider integer spin in this work, since only the integer-spin field can contribute to tree-level inflaton correlators. For the mass $m$, as mentioned at the end of the introduction, we shall always use the \emph{mass parameter} $\wt\nu$ in place of the mass $m$. The two are related by $\wt\nu=\sqrt{m^2-9/4}$ for a scalar field and $\wt\nu=\sqrt{m^2-(s-1/2)^2}$ for a field of spin $s\neq 0$. For simplicity, in the calculations performed in this work, we always assume $\wt\nu$ to be positive real. This is the situation most relevant to CC physics. However, our final analytic results are also applicable to lighter particles by analytic continuation, in which cases the mass parameter $\wt\nu$ takes purely imaginary value.

It is useful to present some basic results about a massive scalar particle. Given a massive scalar field $\si(\tau,\mb x)$ with mass parameter $\wt\nu>0$, we can canonically quantize it in the usual way, by writing it as a linear combination of the creation operator $a_{-\mb k}^\dag$ and annihilation operator $a_{\mb k}$ in the 3-momentum space:
\begin{align}
  \si(\tau,\mb x)=\int\FR{\di^3\mb k}{(2\pi)^3}\Big[\si(k,\tau)a_{\mb k}+\si^*(k,\tau)a^\dag_{-\mb k}\Big]e^{\ii\mb k\cdot\mb x},
\end{align}
where the time-dependent coefficient $\si(k,\tau)$ is called the mode function, and is determined by the Klein-Gordon equation in dS together with the Bunch-Davies initial condition. Explicitly:
\bge
  \si(k,\tau)=\FR{\sqrt\pi}{2}e^{-\pi\wt\nu/2}(-\tau)^{3/2}\text{H}_{\ii\wt\nu}^{(1)}(-k\tau),
\ede
where $\mathrm{H}_{\nu}^{(1)}(z)$ is the Hankel function of first kind. Using the mode function, one can construct the bulk propagator $D_{\aa\bb}(k;\tau_1,\tau_2)$ for $\si$ field, in which $\aa,\bb=\pm$ and $\tau_{1,2}$ are SK indices and conformal time variables at the two bulk endpoints of the propagator. According to the standard procedure in the SK formalism, as reviewed in \cite{Chen:2017ryl}, one first constructs the two Wightman functions $D_>(k;\tau_1,\tau_2)$ and $D_<(k;\tau_1,\tau_2)$. The ``greater'' Wightman function $D_>$ is given, in terms of the mode function $\si(k,\tau)$, by:
\bge
\label{eq_DScalarGreater}
  D_>(k;\tau_1,\tau_2)=\si(k,\tau_1)\si^*(k,\tau_2)=\FR{\pi e^{-\pi\wt\nu}}{4}(\tau_1\tau_2)^{3/2}\mathrm{H}_{\ii\wt\nu}^{(1)}(-k\tau_1)\mathrm{H}_{-\ii\wt\nu}^{(2)}(-k\tau_2),
\ede
while the ``less'' Wightman function $D_<(k;\tau_1,\tau_2)=D_>^*(k;\tau_1,\tau_2)$. Then, the four SK propagators $D_{\aa\bb}$ with $\aa,\bb=\pm$ are given as:
\begin{align}
\label{eq_DScalarSame}
  &D_{\pm\pm}(k;\tau_1,\tau_2)=D_{\gtrless}(k;\tau_1,\tau_2)\theta(\tau_1-\tau_2)+D_\lessgtr(k;\tau_1,\tau_2)\theta(\tau_2-\tau_1),\\ 
\label{eq_DScalarOpp}
  &D_{\pm\mp}(k;\tau_1,\tau_2)=D_\lessgtr(k;\tau_1,\tau_2). 
\end{align}

On the other hand, it is also important to consider the boost-breaking dispersion relations for the massive particle, as mentioned above. In this case, one needs more parameters to characterize the particle. If we break the three dS boost symmetries but retain the dS dilatation, then there are two additional parameters one can include in the quadratic Lagrangian of a massive field: First, one can have a non-unit sound speed $c_s$ which characterizes the relative sizes between the kinetic energy $\si'^2$ and the gradient energy $(\pd_i\si)^2$. The non-unit sound speed can be introduced in our computation by replacing all momentum $k$ of the massive field by $c_s k$. The problem of including non-unit sound speed has been treated in detail in \cite{Pimentel:2022fsc,Jazayeri:2022kjy}, and we do not consider them from now on. 

Second, when the dS boost is broken by, say, a rolling inflaton background, the helicity of a massive spinning particle $h\equiv\mb s\cdot\mb k/|\mb k|$ becomes unambiguously defined, since we are no longer allowed to boost particles into different helicity states. Therefore, one can include a helicity-weighted particle number operator in the Hamiltonian, whose coefficient is a helicity-dependent chemical potential $\wt\mu$ \cite{Wang:2019gbi}, which we shall simply call chemical potential for short.\footnote{The chemical potential $\mu$ is a dim-1 parameter, and it is usually convenient to define a dimensionless chemical potential $\wt\mu=\mu/H$. Since we take $H=1$ in this work, the two become the same. } Technically, the chemical potential modifies the mode function of a massive spinning particle from the original Hankel-type functions to the Whittaker W functions. For example, a massive spin-1 particle of mass parameter $\wt\nu$ and chemical potential $\wt\mu$ has the following mode function $B^{(h)}(k,\tau)$ for its two transverse polarizations $h=\pm 1$:
\bge
\label{eq_Bhmode}
  B^{(h)}(k,\tau)=\FR{e^{-h\pi\wt\mu/2}}{\sqrt{2k}}\mathrm{W}_{\ii h\wt\mu,\ii\wt\nu}(2\ii k\tau),
\ede
where $\text{W}_{\ka,\nu}(z)$ is the Whittaker W function. The bulk propagators for particles of nonzero chemical potential can then be constructed from their mode functions in the same way as shown above for a massive scalar. It is then clear that the cases with nonzero chemical potential require a separate treatment. Therefore, we classify the correlators considered in this work into two categories, one corresponds to the Hankel-type correlators, and includes all massive fields with dS covariant dispersion; the other is the Whittaker-type correlators, which covers the cases of massive spinning fields with nonzero chemical potential. 

With all bulk-to-boundary propagators and bulk propagators specified in Fig.\ \ref{fd_3pt2pt}, it remains to include the appropriate couplings for all the internal vertices. In this work, we allow the massive field $\si$ to have very general couplings to the external inflaton lines. The couplings can be either nonderivative or derivative, and they can contain an arbitrary number of uncontracted time derivatives or contracted spatial derivatives. Normally, if one imposes scale invariance, the time dependence in the coupling coefficient is fully determined by the number of fields and derivatives. However, we do not have to make the assumption of scale invariance, which is broken at least by the inflaton potential. Indeed, it is even possible to include complicated time dependences in the coupling coefficients, such as a component oscillatory in physical time, $c(\tau)\supset (-\tau)^{p}\cos\omega t\sim  (-\tau)^{p\pm\ii\omega}$, which may appear in some resonant models.

\begin{figure}
\centering
  \parbox{0.26\textwidth}{\includegraphics[width=0.26\textwidth]{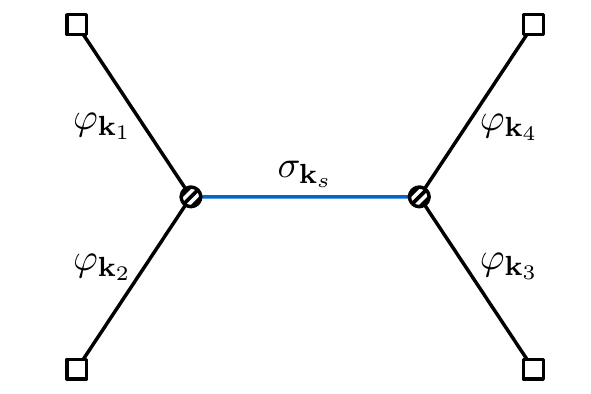}}
  $\xrightarrow[(u_2\to 1)]{\mb k_4\to 0}$
  \parbox{0.26\textwidth}{\includegraphics[width=0.26\textwidth]{fd_scalar_tree_3pt}} 
  $\xrightarrow[(u_1\to 1)]{\mb k_2\to 0}$
  \parbox{0.26\textwidth}{\includegraphics[width=0.26\textwidth]{fd_scalar_tree_2pt}}
  \caption{Three-point and two-point correlators as successive folded limits of a four-point correlator.}
  \label{fd_4pt}
\end{figure}

\paragraph{Seed integrals: definitions and results.} With all ingredients introduced above, we are now in a position to construct correlators and to compute them. The general strategy of this work is that we compute two types of \emph{seed integrals}, one for Hankel-type correlators, and one for Whittaker-type correlators. These seed integrals are essentially the SK integrals for the four-point correlators with $s$-channel exchange of a massive field, shown in Fig.\ \ref{fd_4pt}. As four-point functions with $s$-channel mediation, the seed integrals depend on the magnitudes of the four external momenta, $k_i\equiv |\mb k_i|$ $(i=1,2,3,4)$, as well as the magnitudes of the $s$-channel momentum $k_s\equiv |\mb k_s|$, with $\mb k_s\equiv \mb k_1+\mb k_2$. We also use shorthand notations such as $k_{12}\equiv k_1+k_2$, $k_{34}\equiv k_3+k_4$, and $k_{123}\equiv k_1+k_2+k_3$.

We define the seed integrals such that they depend on these momenta only through two independent ratios. A widely used pair of independent momentum ratios are the following:
\begin{align}
  &r_1\equiv\FR{k_s}{k_{12}}, 
 &&r_2\equiv\FR{k_s}{k_{34}}.
\end{align}
However, a key point in this work is that it is more advantageous to use the following two variables $u_{1,2}$ instead of $r_{1,2}$, as was observed in \cite{Qin:2022fbv}:
\begin{align}
  &u_1\equiv\FR{2k_s}{k_{12}+k_s}, 
 &&u_2\equiv\FR{2k_s}{k_{34}+k_s}. 
\end{align}
As we shall see in later sections, it is the use of the $u$-variables that enables us to conveniently take the folded limits of the four-point function, and to get the closed-form analytical formulae for the three-point and two-point functions.

Now we are ready to introduce the seed integrals, to which a large class of correlators can be reduced, as will be shown later in this section. As explained above, we introduce two types of seed integrals. One is the \emph{Hankel seed integral} $\wt{\mathcal{I}}_{\aa\bb}^{p_1p_2}(u_1,u_2)$, which is defined by:
\begin{keyeqn}
\begin{align}
\label{eq_SeedIntH}
  \wt{\mathcal{I}}_{\aa\bb}^{p_1p_2}(u_1,u_2)\equiv (-\aa\bb)k_s^{5+p_{12}}\int_{-\infty}^0\di\tau_1\di\tau_2(-\tau_1)^{p_1}(-\tau_2)^{p_2}e^{\ii\aa k_{12}\tau_1+\ii\bb k_{34}\tau_2}D_{\aa\bb}(k_s;\tau_1,\tau_2).
\end{align}
\end{keyeqn}
Here $\aa,\bb=\pm1$ are SK indices as required by the diagrammatic rule in the SK formalism, and $(p_1,p_2)$ are a pair of numbers that can take most \emph{complex} values, except that we require $\text{Re}~p_{1,2}>-5/2$ to make the seed integral well defined. Furthermore, $D_{\aa\bb}(k;\tau_1,\tau_2)$ is the bulk SK propagator for a massive scalar field with mass parameter $\wt\nu$, whose explicit form has been given in (\ref{eq_DScalarGreater}), (\ref{eq_DScalarSame}), and (\ref{eq_DScalarOpp}). With these explicit expressions, one can see that the seed integral defined above depends on various momenta only through two independent ratios, which we choose to be $u_1$ and $u_2$. 

The Hankel seed integral will be computed explicitly in Sec.\ \ref{sec_hankel} with one or both of $u_1$ and $u_2$ taken to be 1, so that one can directly read the three-point and two-point correlators from them. The result for the three-point limit $\wt{\mathcal{I}}_{\aa\bb}^{p_1p_2}(u_1,1)$ is summarized in (\ref{eq_IPPresult}) and (\ref{eq_IPMresult}), and the result for the two-point limit $\wt{\mathcal{I}}_{\aa\bb}^{p_1p_2}(1,1)$ is given in (\ref{eq_H2ptResultPP}) and (\ref{eq_H2ptResultPM}).

Likewise, we introduce the following \emph{Whittaker seed integral} for the Whittaker-type correlators:
\begin{keyeqn}
\begin{align}
\label{eq_SeedIntW}
  \wt{\mathcal{I}}^{(h)p_1p_2}_{\aa\bb}(u_1,u_2)\equiv (-\aa\bb)k_s^{3+p_{12}}\int_{-\infty}^0\di\tau_1\di\tau_2(-\tau_1)^{p_1}(-\tau_2)^{p_2}e^{\ii\aa k_{12}\tau_1+\ii\bb k_{34}\tau_2}D_{\aa\bb}^{(h)}(k_s;\tau_1,\tau_2).
\end{align}
\end{keyeqn}
Everything here is in parallel with the Hankel seed integral, except that here we are including the bulk SK propagator $D_{\aa\bb}^{(h)}(k_s;\tau_1,\tau_2)$ for the helicity-$h$ $(h=\pm 1)$ component of a massive spin-1 field, constructed from the mode function (\ref{eq_Bhmode}). Therefore, we add a superscript $(h)$ in the seed integral to distinguish it from the Hankel seed integral. The computation of the Whittaker seed integral will be given in Sec.\ \ref{sec_whittaker}. The result for the three-point limit $\wt{\mathcal I}_{\aa\bb}^{(h)p_1p_2}(u,1)$ is given in (\ref{eq_IhPMresult}) and (\ref{eq_IhPPresult}), and the result for the two-point limit $\wt{\mathcal I}_{\aa\bb}^{(h)p_1p_2}(1,1)$ is given in (\ref{eq_Ih2ptResultPM}) and (\ref{eq_Ih2ptResultPP}).

One notable difference in our definition of seed integrals from the previously defined seed integrals (including our earlier work \cite{Qin:2022fbv}) is that the SK indices $\aa,\bb$ are left unsummed. As we shall see later, when writing correlators as linear combinations of seed integrals, the coefficients of the linear combination usually involve additional SK indices, and this fact necessitates the results of the seed integrals for each individual SK branch. In previous works, this complication was usually overcome by acting appropriate differential operators on a fully summed seed integral. Acting on such differential operators has the effect of raising the powers $p_{1,2}$ by integer units. However, when we want to change $p_{1,2}$ continuously on their complex planes, the method of acting differential operators no longer works. Also, for the three-point and two-point functions, there is simply no sufficient momentum dependence to act the differential operator. For example, to raise the power of $p_2$ in a four-point seed integral, one needs to act a differential operator that contains $\pd_{u_2}$. However, for the three-point function, $u_2$ has been set to a constant, so the differential operator is no longer applicable. It is for these two reasons that we choose to keep the SK indices explicit in the definition of the seed integrals. We shall explain this point more clearly with an example below.

\subsection{Examples}

In this subsection, we provide various examples, showing how to reduce the correlators to seed integrals. We choose a range of examples that often appear in models of CC physics. We also provide sufficient details and mid-steps to illustrate the procedure. 

\subsubsection{Scalar exchange}

The first group of examples all involve tree-level exchange of a massive scalar field $\si$, but the massive scalar can couple to the external inflaton $\varphi$ in many different ways. The first three examples below consider three special choices of couplings, all coming from CC model buildings. The fourth example considers the most general coupling one can write down between $\si$ and $\varphi$.

\paragraph{Time-derivative coupling.} The presence of a rolling inflaton background breaks the time diffeomorphism, and thus allows us to introduce uncontracted temporal indices when writing the effective Lagrangian for the fluctuating fields. In particular, one can consider couplings between $\si$ and the conformal time derivative of $\varphi'\equiv \di\varphi/\di\tau$. Therefore, for the trilinear and bilinear couplings shown in Fig.\ \ref{fd_3pt2pt}, we can choose:
\begin{align}
  &\mathcal{O}_{\varphi\varphi\si}= \FR{1}{2}(-\tau)^{-2}(\varphi')^2\si, 
  &&\mathcal{O}_{\varphi\si}= (-\tau)^{-3}\varphi'\si.
\end{align}
The explicit $\tau$ dependence in the couplings is dictated by the scale invariance, and we do not include the constant coefficients for clarity. It is trivial to put them back in the final expression. The above couplings are perhaps the simplest choice from a purely technical point of view, owing to the fact that the bulk-to-boundary propagator $G_{\aa}(k,\tau)$ in (\ref{eq_Gbtob}) becomes simpler when acting on a conformal time derivative: $G_{\aa}'(k,\tau)=\tau e^{+\ii\aa k\tau}/(2k)$. The SK integral thus takes the following form: (See \cite{Chen:2017ryl} for the diagrammatic rules in the SK formalism.)
\begin{align}
  \la\varphi_{\mb k_1}\varphi_{\mb k_2}\varphi_{\mb k_3}\ra_{\si_{\mb k_3}}'=&-\sum_{\aa,\bb=\pm}\aa\bb\int\FR{\di\tau_1}{(-\tau_1)^2}\FR{\di\tau_2}{(-\tau_2)^3}  G_\aa'(k_1;\tau_1)G_\aa'(k_2;\tau_1)  G_\bb'(k_3;\tau_2) D_{\aa\bb}(k_3;\tau_1,\tau_2)\n\\ 
  =&~\FR{1}{8 k_1k_2k_3}\sum_{\aa,\bb=\pm}\aa\bb\int \di\tau_1 \FR{\di\tau_2}{(-\tau_2)^2} e^{\ii\aa k_{12}\tau_1+\ii\bb k_3\tau_2}D_{\aa\bb}(k_3;\tau_1,\tau_2) .
\end{align}
Here the notation $\la\varphi_{\mb k_1}\varphi_{\mb k_2}\varphi_{\mb k_3}\ra_{\si_{\mb k_3}}'$ means that we have only included the graph with the massive propagator carrying the momentum $\mb k_3$. There are also other two permutations to be included in the full result $\la\varphi_{\mb k_1}\varphi_{\mb k_2}\varphi_{\mb k_3}\ra_\si'$. Now, comparing the above expression with the Hankel seed integral defined in (\ref{eq_SeedIntH}), we immediately find: 
\begin{align}
  \la\varphi_{\mb k_1}\varphi_{\mb k_2}\varphi_{\mb k_3}\ra_\si'
  = \FR{-1}{8k_1k_2k_3^4}\sum_{\aa,\bb=\pm}\wt{\mathcal I}_{\aa\bb}^{0,-2}(u,1)+(\text{2 perms}),
\end{align}
where $u=2k_3/k_{123}$. In a similar way, one can compute the correction to the two-point function $\de\la\varphi_{\mb k}\varphi_{-\mb k}\ra_\si'$ with two insertions of bilinear couplings $\mathcal{O}_{\varphi\si}$ introduced above:
\begin{align}
  \de\la\varphi_{\mb k}\varphi_{-\mb k}\ra_\si'=&-\sum_{\aa,\bb=\pm}\aa\bb\int\FR{\di\tau_1}{(-\tau_1)^3}\FR{\di\tau_2}{(-\tau_2)^3}  G_\aa'(k;\tau_1)G_\bb'(k;\tau_2) D_{\aa\bb}(k;\tau_1,\tau_2)\n\\
  =&~\FR{1}{4k^3}\sum_{\aa,\bb=\pm}\wt{\mathcal I}_{\aa\bb}^{-2,-2}(1,1).
\end{align}
In this simple example, we have only one seed integral with its SK indices directly summed, thanks to the simple form of the bulk-to-boundary propagator with one time derivative. We will see more complicated combinations in the following examples where the combination coefficients themselves contain SK indices.

\paragraph{dS-invariant coupling.} 
Our second example is a slight modification of the first example. Instead of time-derivative coupling for the trilinear coupling, we make the following dS invariant choice:
\begin{align}
  &\mathcal{O}_{\varphi\varphi\si}= \FR{1}{2}(-\tau)^{-2}\eta^{\mu\nu}(\pd_\mu\varphi)(\pd_\nu\varphi)\si, 
  &&\mathcal{O}_{\varphi\si}= (-\tau)^{-3}\varphi'\si.
\end{align}
At the same time, we still keep the time derivative coupling for the bilinear vertex. We note in passing that dS invariant couplings for the bilinear vertex, such as $\pd_\mu\si\pd^\mu\varphi$ and $\si\varphi$, are in a sense trivial, since such couplings can always be rotated away by a simple linear field redefinition. Consequently, dS invariant two-point mixings normally do not lead to  physically interesting effects in CC physics. 

It is straightforward to find the following SK integral for the three-point correlator constructed with the above two couplings: 
\begin{align}
\la\varphi_{\mb k_1}\varphi_{\mb k_2}\varphi_{\mb k_3}\ra_{\si_{\mb k_3}}'
  =&~\sum_{\aa,\bb=\pm}\aa\bb\int\FR{\di\tau_1}{(-\tau_1)^2}\FR{\di\tau_2}{(-\tau_2)^3}  G_\bb'(k_3;\tau_2) D_{\aa\bb}(k_3;\tau_1,\tau_2)\n\\
  &~\times \Big[G_\aa'(k_1;\tau_1)G_\aa'(k_2;\tau_1)+\mb k_1\cdot\mb k_2G_\aa(k_1;\tau_1)G_\aa(k_2;\tau_1)\Big].
\end{align}
Using the explicit expression for the bulk-to-boundary propagator $G_\aa(k;\tau)$ in (\ref{eq_Gbtob}) and then comparing the result with the definition of the seed integral (\ref{eq_SeedIntH}), we have:
\begin{align}
\label{eq_3ptLorCovCoup}
\la\varphi_{\mb k_1}\varphi_{\mb k_2}\varphi_{\mb k_3}\ra_\si'
  =&~\FR{1}{16(k_1k_2k_3)^2}\sum_{\aa,\bb=\pm} \Big[(\varrho_{12}^2-1)\wt{\mathcal I}_{\aa\bb}^{0,-2}(u,1)+\ii\aa \FR{\varrho_{12}(1-\varrho_1^2-\varrho_2^2)}{\varrho_1\varrho_2}\wt{\mathcal I}_{\aa\bb}^{-1,-2}(u,1)\n\\
  &~+\FR{1-\varrho_1^2-\varrho_2^2}{\varrho_1\varrho_2}\wt{\mathcal I}_{\aa\bb}^{-2,-2}(u,1)\Big]+(\text{2 perms}),
\end{align}
where $u=2k_3/k_{123}$ as before, and we have defined $\varrho_1\equiv k_1/k_3$, $\varrho_2\equiv k_2/k_3$, $\varrho_{12}\equiv k_{12}/k_3$ for simplicity. Also, we have used the momentum conservation to write $\mb k_1\cdot\mb k_2=\frac{1}{2}(k_3^2-k_1^2-k_2^2)$. As mentioned before, the coefficients in front of some seed integrals have explicit dependence on SK indices. In this particular example, one can get rid of the SK-indices-dependent coefficients by acting appropriate differential operators. One can directly check that the following equation holds: 
\begin{align}
  \la\varphi_{\mb k_1}\varphi_{\mb k_2}\varphi_{\mb k_3}\ra_\si'
  =&~\FR{1}{8(k_1k_2k_3)^2}\Big[-k_1k_2\pd_{k_{12}}^2+(\wh{\mb k}_1\cdot\wh{\mb k}_2)(1-k_1\pd_{k_{12}})(1-k_2\pd_{k_{12}})\Big]\sum_{\aa,\bb=\pm} \wt{\mathcal I}_{\aa\bb}^{-2,-2}(u,1)\n\\
  &~+(\text{2 perms}).
\end{align}
When acting the differential operator $\pd_{k_{12}}$, the $u$ parameter in $\wt{\mathcal I}_{\aa\bb}^{-2,-2}(u,1)$ should be understood as $u=2k_3/(k_{12}+k_3)$. Thus we see that it is sometimes possible to avoid SK-index-dependent coefficients by acting appropriate differential operators on the fully summed seed integrals. But this is not always possible, as we shall see in a later example with the most general couplings.

\paragraph{Oscillating coupling.} 
Now we consider an example with explicitly time-dependent couplings. In particular, we consider a trilinear coupling that is oscillatory in \emph{physical} time $t$. 
\begin{align}
  &\mathcal{O}_{\varphi\varphi\si}= \FR{1}{2}(-\tau)^{-2}\big[1+\lam(-\tau)^p\cos(\omega t+\de)\big](\varphi')^2\si, 
  &&\mathcal{O}_{\varphi\si}= (-\tau)^{-3}\varphi'\si.
\end{align}
Here $\lam$ is a small dimensionless parameter, $\de$ is a phase, and the real exponent $p$ describes how the oscillation amplitude changes with time. Such a trilinear coupling can appear when there is a background field oscillating with a physical frequency $\omega$.

Now let us focus on the three-point correlator contributed by the oscillatory term in $\mathcal{O}_{\varphi\varphi\si}$, namely, the term proportional to $\lam$. With the same procedure, we write down the SK integral and compare it with the Hankel seed integral, which shows that the $\order{\lam}$ part of the correlator can be written in the following way:
\begin{align}
  \la\varphi_{\mb k_1}\varphi_{\mb k_2}\varphi_{\mb k_3}\ra_\lam' =&~\FR{-\lam}{16k_1k_2k_3^4}\sum_{\aa,\bb=\pm}\Big[e^{-\ii\de}\wt{\mathcal I}_{\aa\bb}^{\,p+\ii\omega,-2}(u,1)+e^{+\ii\de}\wt{\mathcal I}_{\aa\bb}^{\,p-\ii\omega,-2}(u,1)\Big]+(\text{2 perms}).
\end{align}
Here we have used the fact that the physical time is related to the conformal time via $\tau=-e^{-t}$. 

\paragraph{General couplings.}
Finally we consider the most general coupling between the massive scalar field $\si$ and the external inflaton mode $\varphi$ respecting the spatial translation and rotation. Using integration by parts, we can move all the derivatives on $\si$ to $\varphi$. After doing this, the most general trilinear and bilinear couplings take the following form:
\begin{align}
  \mathcal{O}_{\varphi\varphi\si}
  =&~\FR{1}{2}(-\tau)^R \Big[\pd_{\tau}^{J} \pd_{i_1}\cdots\pd_{i_{M}}(\pd_j\pd^j)^{N}\varphi\Big] \Big[\pd_{\tau}^{K}\pd^{i_1}\cdots\pd^{i_{M}}(\pd_k\pd^k)^{L}\varphi\Big] \si,\n\\
  \mathcal{O}_{\varphi\si}
  =&~ (-\tau)^S \Big[\pd_{\tau}^{P}(\pd_i\pd^i)^{Q}\varphi\Big] \si.
\end{align}
Here $R$ and $S$ can be any complex numbers with $\text{Re}\,R,S>-5/2$, while $J,K,M,N,L,P,Q$ can take any nonnegative integer values. If we further assume scale symmetry, then the powers of time $R$ and $S$ are fixed to be $R=-4+J+K+2(M+N+L)$ and $S=-4+P+2Q$, but we do not have to make this assumption. Then, applying the diagrammatic rule in Schwinger-Keldysh formalism, one can write down the SK integral for the three-point correlator:
\begin{align}
&\la\varphi_{\mb k_1}\varphi_{\mb k_2}\varphi_{\mb k_3}\ra_{\si_{\mb k_3}}'
  =-\FR{1}{2}(-\mb k_1\cdot\mb k_2)^M (-k_1^2)^{N}(-k_2^2)^{L}(-k_3^2)^Q\sum_{\aa,\bb=\pm}\aa\bb\int \di\tau_1\di\tau_2(-\tau_1)^R(-\tau_2)^S\n\\
  &~\times\Big[\pd_{\tau_1}^{J}G_\aa(k_1;\tau_1)\Big]\Big[\pd_{\tau_1}^{K}G_\aa(k_2;\tau_1)\Big]\Big[\pd_{\tau_2}^{P}G_\bb(k_3;\tau_2)\Big]D_{\aa\bb}(k_s;\tau_1,\tau_2)+(\mb k_1\leftrightarrow \mb k_2).
\end{align}
To compare this with the Hankel seed integral (\ref{eq_SeedIntH}), it is useful to note the following expression for the bulk-to-boundary propagator with $J$ time derivatives:
\bge
  \pd_{\tau}^{J}G_\aa(k_1;\tau_1)=\FR{1}{2k^3}(\ii\aa k)^J(1-J-\ii\aa k\tau)e^{+\ii \aa k\tau}.
\ede 
Then it is straightforward to find the following expression: 
\begin{align}
&~\la\varphi_{\mb k_1}\varphi_{\mb k_2}\varphi_{\mb k_3}\ra_{\si}'
  =\FR{(-1)^{M+N+L+Q}(\mb k_1\cdot\mb k_2)^M k_1^{2N+J-1} k_2^{2L+K-1}k_3^{2Q+P-R-S-6}}{16k_1^2k_2^2k_3^2}\sum_{\aa,\bb=\pm}(\ii\aa)^J(\ii\aa)^K(\ii\bb)^P  \n\\
   &~\times\bigg[(1-J)(1-K)(1-P)\wt{\mathcal I}_{\aa\bb}^{R,S}(u,1)-\ii\aa\big((1-K)\varrho_1+  (1-J)\varrho_2\big)(1-P)\wt{\mathcal I}_{\aa\bb}^{R+1,S}(u,1)\n\\
   &~-(1-P)\varrho_1\varrho_2\wt{\mathcal I}_{\aa\bb}^{R+2,S}(u,1)-\ii\bb (1-J)(1-K)\wt{\mathcal I}_{\aa\bb}^{R,S+1}(u,1)\n\\
   &~-\aa\bb\big((1-K)\varrho_1+(1-J)\varrho_2\big)\wt{\mathcal I}_{\aa\bb}^{R+1,S+1}(u,1)-\ii\bb \varrho_1\varrho_2\wt{\mathcal I}_{\aa\bb}^{R+2,S+1}(u,1)\bigg]+(\text{5 perms}).
\end{align}
Again, this expression contains coefficients that depend on SK indices. In this example, it is no longer possible to eliminate this dependence by acting on differential operators on a summed seed integral, in particular because the $u_2$ variable has been set to 1. Therefore it seems most convenient to leave the SK indices unsummed in the definition of the seed integral, in order to accommodate the most general couplings as presented here.

\subsubsection{Spinning exchange}

Next we consider the exchange of massive particles with nonzero spin $s$. The angular momentum conservation at each of the two-point mixing vertices in Fig.\ \ref{fd_3pt2pt} implies that only the longitudinal component of the spinning field can mix with the external scalar mode, and that only the helicity-2 component of the spinning field can mix with the external graviton. Therefore, in the following examples, we will only consider longitudinal polarizations for the three-point and two-point scalar correlators with spinning exchange. In the last example, we shall also consider an example of graviton correlators, in which we will have a helicity-2 exchange, possibly with a nonzero chemical potential. 

The main purpose of the following examples is to show how to reduce the SK integrals with spinning field into seed integrals defined with spin-0 or spin-1 massive propagators. Therefore we will no longer consider the most general couplings. Instead, we will focus on the problems such as how to relate the longitudinal mode function of a spinning particle with the mode function of a massive scalar field. We shall begin with the simplest example, namely the spin-1 case, and then consider the general spin-$s$ field. 

\paragraph{Spin-1.}
The three-point function with longitudinal spin-1 exchange has been extensively studied in \cite{Qin:2022fbv}, and here we show the main results together with some important mid-steps for illustration. A massive spin-1 particle can be obtained after canonically quantizing a vector field $A_\mu$ with mass parameter $\wt\nu$. For this field, we choose the following simple couplings to the external inflaton modes: 
\begin{align}
  &\mathcal{O}_{\varphi\varphi A}=\FR{1}{2}(-\tau)^{-1}\varphi'(\pd_i\varphi)A_i,
 &&\mathcal{O}_{\varphi A}=(-\tau)^{-1}(\pd_i\varphi')A_i 
\end{align}
Then, the SK integral for the three-point function mediated by the longitudinal mode of $A$ reads:
\begin{align}
\label{eq_SKintA1}
  \la\varphi_{\mb k_1}\varphi_{\mb k_2}\varphi_{\mb k_3}\ra_{A}'
 =&~\bigg\{ \FR{1}{2}\Big[\mb k_2\cdot\bm{\ep}^{(L)}(\wh k_3)\Big]\Big[\mb k_3\cdot\bm{\ep}^{(L)*}(\wh k_3)\Big]\sum_{\aa,\bb=\pm}\aa\bb\int\FR{\di\tau_1}{-\tau_1}\FR{\di\tau_2}{-\tau_2}\n\\
  &~\times G_{\aa}'(k_1,\tau_1)G_{\aa}(k_2,\tau_1)G_{\bb}'(k_3,\tau_2) D^{(L)}_{\aa\bb}(k;\tau_1,\tau_2)\bigg\} +(\text{5 perms}).
\end{align}
Here we have introduced the longitudinal polarization vector $\bm{\ep}^{(L)}(\wh k)=\wh{\mb k}\equiv \mb{k}/k$, as well as the longitudinal bulk propagator $D_{\aa\bb}^{(L)}(k_3;\tau_1,\tau_2)$. This bulk propagator is built from the longitudinal mode function $B^{(L)}(k,\tau)$ in (\ref{eq_Bhmode}) following the standard procedure. Now, by a careful examination of the equation of motion of $A_\mu$ as well as the constraint $\nabla^\mu A_\mu=0$, one can show that the longitudinal mode function $B^{(L)}(k,\tau)$ can be built from a massive scalar mode function of Hankel-type \cite{Lee:2016vti,Qin:2022fbv}. Here we directly quote the result. It is convenient to define the following differential operator $\mathcal{L}_\tau$:
\begin{align}
  \mathcal{L}_\tau= \pd_\tau-\FR{2}{\tau} .
\end{align}
Then, one can show that the longitudinal component in $A_i$ is related to the temporal polarization $B^{(L)}_0$ in $A_0$ via an action of $\mathcal{L}_\tau$ operator. In addition, the mode function for the temporal polarization $B^{(L)}_0$ is identical to that of the massive scalar $\si(k,\tau)$ up to a constant factor: 
\begin{align}
  &B^{(L)}(k,\tau)=\FR{1}{\ii k}\mathcal{L}_\tau B_0^{(L)}(k,\tau),\\
  &B^{(L)}_0(k,\tau)=\FR{\sqrt\pi k}{2m}e^{-\pi\wt\nu/2}(-\tau)^{3/2}\mathrm{H}^{(1)}_{\ii\wt\nu}(-k\tau)=\FR{k}{m}\si(k,\tau). 
\end{align}
From the above relations, we can immediately find a relation between the longitudinal propagator $D_{\aa\bb}^{(L)}(k;\tau_1,\tau_2)$ and the massive scalar propagator $D_{\aa\bb}(k;\tau_1,\tau_2)$:
\begin{align}
\label{eq_DLtoD}
  D_{\aa\bb}^{(L)}(k;\tau_1,\tau_2)=\FR{1}{m^2}\Big[\mathcal{L}_{\tau_1}\mathcal{L}_{\tau_2}D_{\aa\bb}(k;\tau_1,\tau_2)-\ii\aa\tau_1\tau_2\de_{\aa\bb} \de(\tau_1-\tau_2)\Big].
\end{align}
There is a notable contact term proportional to $\de_{\aa\bb}\de(\tau_1-\tau_2)$ in this expression. It is nonzero only for the two same-sign propagators $D_{\pm\pm}^{(L)}$. This term arises when we commute the $\mathcal{L}_\tau$ operator, originally acting on the mode function, with the step function in $D_{\pm\pm}^{(L)}$; See (\ref{eq_DScalarSame}).

Now we substitute (\ref{eq_DLtoD}) in (\ref{eq_SKintA1}), which then leads to the following SK integral involving the massive scalar propagator $D_{\aa\bb}(k;\tau_1,\tau_2)$:
\begin{align}
 \la\varphi_{\mb k_1}\varphi_{\mb k_2}\varphi_{\mb k_3}\ra_{A}'
 =&~\FR{\mb k_2\cdot\mb k_3}{2m^2}\bigg\{\sum_{\aa,\bb=\pm}\aa\bb\int\FR{\di\tau_1}{-\tau_1}\FR{\di\tau_2}{-\tau_2}G_{\aa}'(k_1;\tau_1)G_{\aa}(k_2;\tau_1)G_{\bb}'(k_3;\tau_2)\mathcal{L}_{\tau_1}\mathcal{L}_{\tau_2}D_{\aa\bb}(k;\tau_1,\tau_2)\n\\
  &-\sum_{\aa=\pm}\ii \aa\int \di\tau_1\,G_{\aa}'(k_1;\tau_1)G_{\aa}(k_2;\tau_1)G_{\aa}'(k_3;\tau_1)\bigg\}+(\text{5 perms}).
\end{align}
The first term in the second line comes from the contact term in (\ref{eq_DLtoD}). Then, we perform integration by parts, to move $\mathcal{L}_{\tau_1}$ and $\mathcal{L}_{\tau_2}$ to external modes. There is no boundary term since the integrand vanishes at both $\tau=-\infty$ (by $\ii\ep$ prescription enforced by the Bunch-Davies initial condition) and $\tau=0$ (by explicitly counting the powers of $\tau_{1,2}$ in the integrand). After this is done, we are ready to write the above SK integral as a linear combination of the Hankel seed integral: 
\begin{align}
  \la\varphi_{\mb k_1}\varphi_{\mb k_2}\varphi_{\mb k_3}\ra_{A}'
  =&-\FR{1}{m^2}\FR{\wh{\mb k}_2\cdot\wh{\mb k}_3}{16(k_1k_2k_3)^2}\bigg\{\sum_{\aa,\bb=\pm} \bigg[4\varrho_1\wt{\mathcal I}_{\aa\bb}^{-1,-1}(u,1)-2\ii\bb\varrho_1\wt{\mathcal I}_{\aa\bb}^{-1,0}(u,1)\n\\
  &-2\ii\aa\varrho_1(\varrho_1-2\varrho_2)\wt{\mathcal I}_{\aa\bb}^{0,-1}(u,1)-\aa\bb\varrho_1(\varrho_1-2\varrho_2)\wt{\mathcal I}_{\aa\bb}^{0,0}(u,1)+2\varrho_1\varrho_2\varrho_{12}\wt{\mathcal I}_{\aa\bb}^{1,-1}(u,1)\n\\
  &-\ii\bb \varrho_1\varrho_2\varrho_{12}\wt{\mathcal I}_{\aa\bb}^{1,0}(u,1)\bigg]+\FR{4\varrho_1\varrho_3^2(\varrho_{123}+3\varrho_2)}{\varrho_{123}^4}\bigg\}+(\text{5 perms}).
\end{align} 
Here again $u=2k_3/k_{123}$, and we define the same momentum ratios $\varrho_i$ as in (\ref{eq_3ptLorCovCoup}). Therefore, we see that the computation of three-point correlators with spin-1 exchange can also be reduced to that of the Hankel seed integral.

\paragraph{General integer spin.} 
The procedure described above for spin-1 field can be readily applied to fields with general integer spin $s$, although the algebra can be quite tedious. A spin-$s$ particle can be described by a transverse, traceless, and totally symmetric tensor $\Psi_{\mu_1\cdots\mu_s}$ of rank-$s$. 
There are again a lot of possible choices for the trilinear and bilinear vertices in Fig.\ \ref{fd_3pt2pt}. We do not pursue the most general possibilities, but only illustrate the method with the following simple example:
\begin{align}
  &\mathcal{O}_{\varphi\varphi\si}=(-\tau)^{-3+2s}\varphi'(\pd_{i_1}\cdots \pd_{i_s}\varphi)\Psi_{i_1\cdots i_s},
  &&\mathcal{O}_{\varphi\si}=(-\tau)^{-3+2s}(\pd_{i_1}\cdots\pd_{i_s}\varphi')\Psi_{i_1\cdots i_s},
\end{align}
where the explicit time dependence is again fixed by the scale symmetry. With the couplings given, the SK integral for the three-point correlator with $\Psi$-exchange reads:
\begin{align}
  \la\varphi_{\mb k_1}\varphi_{\mb k_2}\varphi_{\mb k_3}\ra_{\Psi}'
 =&~\bigg\{\FR{(-1)^{s+1}}{2}\Big[k_2^{i_1}\cdots k_2^{i_s}{\ep}_{i_1\cdots i_s}^{(L)}({\wh k}_3)\Big]\Big[ k_3^{j_1}\cdots k_3^{j_s}{\ep}_{j_1\cdots j_s}^{(L)*}({\wh k}_3)\Big]\sum_{\aa,\bb=\pm}\aa\bb\int\FR{\di\tau_1}{(-\tau_1)^{3-2s}}\FR{\di\tau_2}{(-\tau_2)^{3-2s}}\n\\
  &~\times G_{\aa}'(k_1,\tau_1)G_{\aa}(k_2,\tau_1)G_{\bb}'(k_3,\tau_2) D^{(s,L)}_{\aa\bb}(k;\tau_1,\tau_2)\bigg\} +(\text{5 perms}).
\end{align}
Here $\ep_{i_1\cdots i_s}^{(L)}(\wh{k})$ is the longitudinal polarization tensor. Up to a normalization factor, $\ep_{i_1\cdots i_s}^{(L)}(\wh{k})$ is the unique combination of $\wh{k}_i$ and $\de_{ij}$ with the property of being transverse, traceless, and totally symmetric with all indices. Therefore, in the above expression, the factor $k_3^{j_1}\cdots k_3^{j_s}{\ep}_{j_1\cdots j_s}^{(L)*}({\wh k}_3)$ contributes a constant factor, while $k_2^{i_1}\cdots k_2^{i_s}{\ep}_{i_1\cdots i_s}^{(L)}({\wh k}_3)$ contributes to a angular dependent factor $\propto \mathrm{P}_{s}(\cos\theta_{23})$ where $\mathrm{P}_n(z)$ is the Legendre polynomial and $\theta_{23}$ is the angle between $\mb k_2$ and $\mb k_3$. The factor $\mathrm{P}_n(z)$ is well known in CC physics as the signal of a spin-$s$ exchange. However, let us emphasize that this factor also depends on the form of the couplings. Had we coupled the spin-$s$ in a different way to the external mode, this factor will change accordingly.

The propagator $D_{\aa\bb}^{(s,L)}$ corresponding to the longitudinal polarization tensor $\ep_{i_1\cdots i_s}^{(L)}$ with $s$ spatial indices is again built from the mode function for the helicity-0 component of the spin-$s$ field with all-spatial indices $\Psi_{i_1\cdots i_s}$. Similar to the previous spin-1 case, the mode function $\Psi^{(s,L)}_s(k,\tau)$ can also be constructed from the scalar mode function $\si(k,\tau)$ of the same mass parameter $\wt\nu$ by the action of a differential operator. For spin-$s$, this differential operator is defined recursively. Here we quote the result, and the details can be found in, e.g., App.\ A of \cite{Lee:2016vti}. First, similar to spin-1 case, the temporal mode function $\Psi^{(s,L)}_0(k,\tau)$ is identical to the scalar mode function up to a constant factor:
\begin{align}
  \Psi^{(s,L)}_0(k,\tau)= \FR{\pi^{3/4}\text{sech}^{1/2}(\pi\wt\nu)}{2^{s/2}}\Gamma^{1/2}\bgb 1+s \\ \fr12+s,\fr12+s+\ii\wt\nu,\fr12+s-\ii\wt\nu\edb  k^s\si(k,\tau),
\end{align}
where we have used the compact notation for the Euler-$\Gamma$ product, as defined in (\ref{eq_GammaProd2}). Then, the longitudinal mode function $\Psi^{(s,L)}_s(k,\tau)$ with respect to the longitudinal polarization of $s$ spatial indices is constructed recursively from the following relation:
\begin{align}
  \Psi^{(s,L)}_{n}(k,\tau)=\FR{1}{\ii k}\mathcal{L}_\tau \Psi^{(s,L)}_{n-1}(k,\tau)-\sqrt\pi\sum_{m=0}^{n-1}\Gamma\bgb 1+n,\fr{1+m+n}2 \\ 1+m,1-m+n,\fr12+n,\fr{1+m-n}2\edb \Psi^{(s,L)}_{m}(k,\tau).
\end{align}
For example, the all-spatial longitudinal mode functions, $\Psi_2^{(2,L)}(k,\tau)$ for a spin-2 particle and $\Psi_3^{(3,L)}(k,\tau)$ for a spin-3 particle, are respectively given by:
\begin{align}
  \Psi_2^{(2,L)}(k,\tau)
  =&~\sqrt{\FR{2}{3(\wt\nu^2+\fr14)(\wt\nu^2+\fr94)}}\bigg(\mathcal{L}_\tau^2-\FR{k^2}{3}\bigg)\si(k,\tau),\\
  \Psi_3^{(3,L)}(k,\tau)
  =&~\sqrt{\FR{2}{5(\wt\nu^2+\fr14)(\wt\nu^2+\fr94)(\wt\nu^2+\fr{25}4)}}\bigg(\mathcal{L}_\tau^3-\FR{14 k^2}{15}\mathcal{L}_\tau\bigg)\si(k,\tau),
\end{align}
where we have dropped unimportant overall phases. Then, in parallel with the previous \mbox{spin-1} example, we can move the differential operator on $\si(k,\tau)$ to the external modes by using the integration by parts, and thereby reduce the correlator of spin-$s$ exchange to a linear combination of the Hankel seed integral.

\paragraph{Tensor and mixed correlators.} In all examples considered above, the correlators are reduced to linear combinations of Hankel seed integral. The Whittaker seed integral has not been used until now. The reason is that the Whittaker mode function appears only for spinning field with helicity-dependent chemical potentials. Such chemical potentials usually vanish for the longitudinal polarization, and therefore it does not play a role in the scalar two-point and three-point correlators at the tree level. (However, the chemical-potential-enhanced mode does play a role in scalar three-point function at the one-loop level or the scalar four-point function at the tree level.) To see the effect of the chemical potential in tree-level three-point functions, we should consider spinning external states, among which the graviton correlators are most relevant to realistic CC physics. Indeed, it was shown in \cite{Tong:2022cdz} that a helical chemical potential can be introduced to a massive spin-2 particle $\Sigma$. When $\Sigma$ linearly mixes with the massless graviton mode $\ga_{\mb k}^{(\pm 2)}$, it can mediate a mixed three-point function $\la\varphi_{\mb k_1}\varphi_{\mb k_2}\ga_{\mb k_3}^{(\pm 2)}\ra'$, described again by the left diagram of Fig.\ \ref{fd_3pt2pt}, but with the right external mode replaced by $\ga_{\mb k_3}$. There is also a two-point function of graviton modes $\la\ga_{\mb k}^{(\pm 2)}\ga_{-\mb k}^{(\pm 2)}\ra'$ contributed by the massive spin-2 mode, similar to the right diagram of Fig.\ \ref{fd_3pt2pt}. The corresponding couplings can be chosen as:
\begin{align}
   &\mathcal{O}_{\varphi\varphi\Sigma}=\FR{1}{2}(\pd_i\varphi)(\pd_j\varphi)\si_{ij},
   &&\mathcal{O}_{\ga\Sigma}=(-\tau)^{-1}\gamma_{ij}'\si_{ij}.
\end{align}
To write down the SK integral, we also need the bulk-to-boundary propagator $T_{\aa}(k;\tau)$ for the external graviton, as well as the mode function $\Sigma^{(\pm 2)}(k,\tau)$ for the helicity-2 component of the massive spin-2 field $\Sigma$. They are respectively given by: 
\begin{align}
  &T_\aa(k,\tau)=\FR{1}{4k^3}(1-\ii\aa k\tau)e^{\ii\aa k\tau},\\
  &\Sigma^{(\pm 2)}(k,\tau) = -\FR{e^{\mp \pi\wt\mu/2}}{2\sqrt{k}\tau}\mathrm{W}_{\pm \ii\wt\mu,\ii\wt\nu}(2\ii k\tau)=-\FR{1}{\sqrt2 \tau}B^{(\pm1)}(k,\tau),
\end{align}
where $B^{(\pm1)}(k,\tau)$ is the mode function for the massive spin-1 field with helicity $h=\pm1$, given in (\ref{eq_Bhmode}). From this one can build the SK integral, and then rewrite it as a linear combination of Whittaker seed integrals. The details have been spelled out in \cite{Qin:2022fbv}. Here we show the final result:
\begin{align} 
  &\la \varphi(\mb k_1)\varphi(\mb k_2)\gamma^{(\pm 2)}(\mb k_3)\ra'\n\\
  =&~\FR{\wh{k}_{1i}\wh{k}_{2j} \ep_{ij}^{(\pm 2)}(\wh k_3) }{32(k_1k_2k_3)^2}\sum_{\aa,\bb=\pm}\Big[\wt{\mathcal I}_{\aa\bb}^{(\pm)-1,-1}(u,1)
   +\ii\aa\varrho_{12}\wt{\mathcal I}_{\aa\bb}^{(\pm)0,-1}(u,1)-\varrho_1\varrho_2\wt{\mathcal I}_{\aa\bb}^{(\pm)1,-1}(u,1)\Big],
\end{align}
where, again, $u=2k_3/k_{123}$, and the various momentum ratios $\varrho_i$ are defined in the same way as in (\ref{eq_3ptLorCovCoup}). It turns out that this correlator can also be obtained by acting an appropriate differential operator on a fully summed SK integral. Interested readers can find more details in \cite{Qin:2022fbv}.

It seems that the Whittaker seed integral is of limited use in representing inflation correlators. As we shall show in a separate work \cite{qin_box}, this seed integral will be more useful when we treat the two-point or three-point functions considered here as subgraphs of more complicated diagrams. In that case, the Whittaker seed integral will be a very convenient building block for constructing more complicated processes at the loop level. 

\section{Bootstrapping Hankel Seed Integrals}
\label{sec_hankel}

In this section, we provide the details of computing the three-point and two-point functions with Hankel-type tree-level mediation. We will derive the bootstrap equations satisfied by the seed integrals, first in $r$-variables, and then in $u$-variables. We then solve the bootstrap equations and obtain both the particular solution to the inhomogeneous equation and the general solution to the homogeneous equation. The final answer is determined by proper boundary conditions, which we choose to impose from the squeezed limit. Then, we take the single and double folded limit, to get the desired result for the seed integrals with one or both of $u_1$ and $u_2$ taken to 1. Readers uninterested in the technical details can directly go to Sec.\ \ref{sec_hankel_summary} for the final results.
 
\subsection{Hankel seed integral and its bootstrap equation}

As mentioned above, our starting point is the Hankel seed integral $\mathcal{I}_{\aa\bb}^{p_1p_2}(u_1,u_2)$ defined in (\ref{eq_SeedIntH}). Here we shall first write the seed integrals as functions of $r_1=k_s/k_{12}$ and $r_2=k_s/k_{34}$, and transform to $(u_1,u_2)$ later:
\begin{align}
\label{eq_seedHr}
\mathcal I^{p_1p_2}_{\aa\bb}(r_1,r_2) 
\equiv - \aa\bb\, k_s^{5+p_{12}} \int_{-\infty}^{0} \di\tau_1\di\tau_2\,(-\tau_1)^{p_1}(-\tau_2)^{p_2}e^{\ii\aa k_{12}\tau_1+\ii \bb k_{34}\tau_2}D_{\aa\bb}(k_s;\tau_1,\tau_2).
\end{align}
where $D_{\aa\bb}(k;\tau_1,\tau_2)$ is the bulk propagator for a massive scalar field, given in (\ref{eq_DScalarSame}) and (\ref{eq_DScalarOpp}). An important and well-known property of the Schwinger-Keldysh propagators is that they satisfy the equation of motion for the massive scalar field with or without a $\delta$-source:
\begin{align}
\label{eq_DEoM1}
&(\tau_1^2 \partial_{\tau_1}^2 - 2\tau_1 \partial_{\tau_1} + k_s^2\tau_1^2 + m^2)D_{\pm\mp}(k_s;\tau_1,\tau_2)=0,\\
\label{eq_DEoM2}
&(\tau_1^2 \partial_{\tau_1}^2 - 2\tau_1 \partial_{\tau_1} + k_s^2\tau_1^2 + m^2)D_{\pm\pm}(k_s;\tau_1,\tau_2)=\mp\ii \tau_1^2\tau_2^2\delta(\tau_1-\tau_2).
\end{align} 
This will be the key property we shall make use of when deriving the bootstrap equation for the Hankel seed integral. 

Now, we start to derive the bootstrap equation. It turns out convenient to redefine several variables. First, we shall use $z_1\equiv -k_{12}\tau_1$ and $z_2\equiv -k_{34}\tau_2$ instead of the conformal time $\tau_1$ and $\tau_2$. Notice that the negative sign in the definition of $z_{1,2}$ makes them real and positive: $z_{1,2}\in(0,+\infty)$ in the physical regime. Second, we shall define a ``hatted'' massive propagator $\wh D_{\aa\bb}(z_1,z_2)$:
\bge
\wh D_{\aa\bb}(z_1,z_2) = k^{3}D_{\aa\bb}(k;\tau_1,\tau_2)=\FR{\pi e^{-\pi\wt\nu}}{4}(z_1z_2)^{3/2}\mathrm{H}_{\ii\wt\nu}^{(1)}(z_1)\mathrm{H}_{-\ii\wt\nu}^{(2)}(z_2).
\ede
The nice thing about the hatted propagator is that it depends on the time and the momentum only through the combination $k\tau_1$ and $k\tau_2$, and therefore, we can rewrite it as a function of two independent variables $z_1$ and $z_2$. With these redefinitions, we can rewrite the Hankel seed integral (\ref{eq_seedHr}) as 
\begin{align}
\label{eq_seedHhatted}
\mathcal I_{\aa\bb}^{p_1p_2}(r_1,r_2)  
=&~(-\aa\bb) r_1^{1+p_1}r_2^{1+p_2} \int_0^\infty \di z_1\di z_2\,z_1^{p_1}z_2^{p_2}e^{-\ii\aa z_1-\ii \bb z_2}\wh D_{\aa\bb}(r_1z_1,r_2z_2).
\end{align}
It now becomes manifest that the Hankel seed integral depends on external momenta only through $r_1$ and $r_2$. 

Recall that the massive propagator $D_{\aa\bb}$ satisfies the equation of motion (\ref{eq_DEoM1}) and (\ref{eq_DEoM2}). This pair of equations can be rewritten in terms of the hatted propagator as
\begin{align}
\label{eq_DhatEoM1}
&(z_1^2 \pd_{z_1}^2 -2 z_1\pd_{z_1}+z_1^2+m^2)\wh D_{\pm\mp}(z_1,z_2) = 0,\\
\label{eq_DhatEoM2}
&(z_1^2 \pd_{z_1}^2 -2 z_1\pd_{z_1}+z_1^2+m^2)\wh D_{\pm\pm}(z_1,z_2) = \mp\ii z_1^2z_2^2\de(z_1-z_2).
\end{align}
For the hatted propagator $\wh D_{\aa\bb}(r_1z_1,r_2z_2)$ in (\ref{eq_seedHhatted}), the two variables $z_1$ and $z_2$ are multiplied by the momentum ratios $r_1$ and $r_2$, respectively. Therefore, for this particular propagator, we can change the variables in the equations (\ref{eq_DhatEoM1}) and (\ref{eq_DhatEoM2}), and rewrite them as differential equations with respect to $r_1$ and $r_2$:
\begin{align}
&(r_1^2 \pd_{r_1}^2 -2 r_1\pd_{r_1}+r_1^2z_1^2+m^2)\wh D_{\pm\mp}(r_1z_1,r_2z_2) = 0,\\
&(r_1^2 \pd_{r_1}^2 -2 r_1\pd_{r_1}+r_1^2z_1^2+m^2)\wh D_{\pm\pm}(r_1z_1,r_2z_2) = \mp\ii r_1^2z_1^2r_2^2z_2^2\de(r_1z_1-r_2z_2),
\end{align}
Now, we can insert the differential operator $(r_1^2 \pd_{r_1}^2 -2 r_1\pd_{r_1}+r_1^2z_1^2+m^2)$ in front of the hatted propagator in (\ref{eq_seedHhatted}), which reduces the massive propagator into 0 or a term proportional to $\delta$ function. Either way, the integral is trivialized, and this will become the right-hand side of the final bootstrap equation. 

To derive the left-hand side of the bootstrap equation, we commute the differential operator $(r_1^2 \pd_{r_1}^2 -2 r_1\pd_{r_1}+r_1^2z_1^2+m^2)$ with the $z$-integrals. It is clear that the only obstacle in our attempt of pulling the differential operator to the left of the integral sign is the $r_1^2z_1^2$ term, which depends on the integration variable $z_1$. This obstacle can be removed by integration by parts. That is, for any well-behaved function $f(rz)$, we have:
\bge
0=\int_0^\infty \di z\, \pd_z\big[z^{p+1} e^{-\ii\aa z} f(rz) \big]=  \int_0^\infty \di z\, z^p e^{-\ii\aa z}( p+1-\ii\aa z+ r\pd_r) f(rz),
\ede
which then gives
\bge
\label{eq_zIBP}
 \int_0^\infty \di z\, z^p e^{-\ii\aa z} zf(rz) = -\ii\aa (r\pd_r+p+1) \int_0^\infty \di z\, z^p e^{-\ii\aa z}f(rz).
\ede
A repetition of the same procedure then gives:
\begin{align}
\label{eq_z2IBP}
 \int_0^\infty \di z\, z^p e^{-\ii\aa z} z^2f(rz) = - (r\pd_r+p+2) (r\pd_r+p+1) \int_0^\infty \di z\, z^p e^{-\ii\aa z}f(rz),
\end{align}
which shows that the $z^2$ term can be as well transformed into a differential operator with respect to $r$. Then, it is straightforward to get the following differential equations:
\begin{align}
&\Big[(r_1^2-r_1^4)\pd_{r_1}^2-\big(2r_1+(4+2p_1)r_1^3\big)\pd_{r_1}
+\big((\wt\nu^2+\fr94)-(p+1)(p+2)r_1^2\big)
\Big]\n\\
&\times \big[r_1^{-1-p_1}r_2^{-1-p_2}\mathcal I_{\pm\mp}^{p_1p_2}(r_1,r_2)\big]=0,\\
&\Big[(r_1^2-r_1^4)\pd_{r_1}^2-\big(2r_1+(4+2p_1)r_1^3\big)\pd_{r_1}
+\big((\wt\nu^2+\fr94)-(p+1)(p+2)r_1^2\big)
\Big]\n\\
&\times \big[r_1^{-1-p_1}r_2^{-1-p_2}\mathcal I_{\pm\pm}^{p_1p_2}(r_1,r_2)\big]=
e^{\mp\ii p_{12}\pi/2}\FR{r_1^{4+p_2}r_2^{4+p_1}}{(r_1+r_2)^{5+p_{12}}}\Gamma(5+p_{12}).
\end{align}
After slight simplifications, we arrived at the bootstrap equations for the Hankel seed integral in terms of $r$-variables: 
\begin{align}
&\mathcal{D}_{r_1}^{p_1}\mathcal I_{\pm\mp}^{p_1p_2}(r_1,r_2)=0,\\
&\mathcal{D}_{r_1}^{p_1}
\mathcal I_{\pm\pm}^{p_1p_2}(r_1,r_2) =
e^{\mp\ii p_{12}\pi/2}\Gamma(5+p_{12})\Big(\FR{r_1r_2}{r_1+r_2}\Big)^{5+p_{12}};\\
&\mathcal{D}_r^p\equiv (r^2-r^4)\pd_{r}^2-\big[(4+2p)r+2r^3\big]\pd_{r}
+ \wt\nu^2+\FR{(5+2p)^2}4.
\end{align}
One can well work with this set of equations and solve for the Hankel seed integral $\mathcal{I}_{\aa\bb}$, as in most previous works on this topic. However, a crucial observation made in \cite{Qin:2022fbv} is that it is much easier to take the three-point and two-point limits if we choose in instead to work with the following new variables:
\bge
u_i \equiv \FR{2r_i}{1+r_i},\quad i=1,2.
\ede
Correspondingly, we rewrite the Hankel seed integral as functions of $u_{1,2}$, namely, we define:
\bge
\wt{\mathcal I}_{\aa\bb}^{p_1p_2}(u_1,u_2)\equiv \mathcal I_{\aa\bb}^{p_1p_2}\big(r_1(u_1),r_2(u_2)\big). 
\ede
Then, the bootstrap equations for $\wt{\mathcal I}_{\aa\bb}^{p_1p_2}(u_1,u_2)$ are:
\begin{align}
\label{eq_BTEs1}
&\mathcal D_{u_1}^{p_1}\wt{\mathcal I}_{\pm\mp}^{p_1p_2}(u_1,u_2)=0,\\
\label{eq_BTEs2}
&\mathcal D_{u_1}^{p_1}\wt{\mathcal I}_{\pm\pm}^{p_1p_2}(u_1,u_2)=
e^{\mp \ii p_{12}\pi/2}\Gamma(5+p_{12})\Big(\FR{u_1u_2}{2(u_1+u_2-u_1u_2)}\Big)^{5+p_{12}};\\
&\mathcal D_u^p \equiv (u^2-u^3)\pd_u^2 - \Big[(4+2p)u-(1+p)u^2\Big]\pd_u + \Big[\wt\nu^2+\big(p+\fr52\big)^2\Big].
\end{align}
This completes the derivation of the bootstrap equations for the Hankel seed integral in $u$-variables. Below we are going to solve these equations to get the four-point function and especially its folded limit.

\subsection{Solving the bootstrap equation}

Now we start to solve the bootstrap equations (\ref{eq_BTEs1}) and (\ref{eq_BTEs2}) for the Hankel seed integrals. 

For the homogeneous equation (\ref{eq_BTEs1}), the solution is a proper linear combination of two independent solutions:
\bge
\label{eq_IpmmpPar}
  \wt{\mathcal{I}}_{\pm\mp}^{p_1p_2}(u_1,u_2)=\sum_{\aa,\bb=\pm}\al_{\pm\mp|\aa\bb}\mathcal{Y}_{\aa}^{p_1}(u_1)\mathcal{Y}_{\bb}^{p_2}(u_2).
\ede
Here $\mathcal{Y}_\pm^p(u)$ is a pair of linearly independent solutions to the homogeneous equation (\ref{eq_BTEs1}), which we choose to be:
\bge
\label{eq_calYpm}
\mathcal Y_\pm^{p_1}(u_1) = 2^{\mp\ii\wt\nu}\Big(\FR{u_1}2\Big)^{5/2+p_1\pm\ii\wt\nu}
\Gamma\Big[\FR52+p_1\pm\ii\wt\nu,\mp\ii\wt\nu\Big]
{}_2\mathrm F_1\left[\bgm\fr52+p_1\pm\ii\wt\nu,\fr12\pm\ii\wt\nu\\1\pm2\ii\wt\nu\edm\middle|u_1\right].
\ede
Here we have included some numerical factors in $\mathcal Y_\pm^{p_1}(u_1)$ for later convenience. Note that we have used the fact that the seed integral $\wt{\mathcal{I}}_{\pm\mp}^{p_1p_2}(u_1,u_2)$ has a symmetry with respect to the exchange $u_1\leftrightarrow u_2$, and that the seed integrals satisfy an identical set of bootstrap equations with respect to $u_2$. Therefore, we expect that the final result for $\wt{\mathcal{I}}_{\pm\mp}^{p_1p_2}(u_1,u_2)$ should be bilinear in $\mathcal{Y}_{\aa}^{p_1}(u_1)$ and $\mathcal{Y}_{\bb}^{p_2}(u_2)$. The coefficients $\al_{\pm\mp|\aa\bb}$ should be determined by imposing a proper boundary condition. In this work, we impose the boundary condition by matching the result to an explicit calculation in the squeezed limit $u_1\ll u_2\ll 1$, as will be elaborated below.

For the inhomogeneous equation (\ref{eq_BTEs2}), the solution is the sum of two parts: one particular solution $\wt{\mathcal{I}}_{\pm\pm,\text{Inh}}^{p_1p_2}(u_1,u_2)$ with the presence of the source term, and one homogeneous solution which is again a linear combination of two independent solutions to the homogeneous equation:
\begin{align}
\label{eq_IpmpmPar}
  \wt{\mathcal{I}}_{\pm\pm}^{p_1p_2}(u_1,u_2)=\wt{\mathcal{I}}_{\pm\pm,\text{Inh}}^{p_1p_2}(u_1,u_2)+\sum_{\aa,\bb=\pm}\al_{\pm\pm|\aa\bb}\mathcal{Y}_{\aa}^{p_1}(u_1)\mathcal{Y}_{\bb}^{p_2}(u_2).
\end{align}
The coefficients $\al_{\pm\pm;\aa\bb}$ are again determined by matching the result to an explicit calculation in the squeezed limit. 

Below, we shall first work out the particular solution $\wt{\mathcal{I}}_{\pm\pm,\text{Inh}}^{p_1p_2}(u_1,u_2)$, and then construct the homogeneous solution. To find the particular solution, we do not work with arbitrary momentum configuration; Instead, we shall directly work with the three-point limit by taking $u_2\to 1$, since we are concerned only with the three-point and two-point functions in this work, and the single folded limit $u_2\to 1$ is regular for the particular solution. The general four-point correlator for arbitrary momentum configuration can be calculated following the same procedure in our previous work \cite{Qin:2022fbv}.

On the other hand, to find the homogeneous solution, we still work with the 4-point function, and impose the boundary conditions in the squeezed limit $u_{1,2}\to 0$. It turns out that this limit is enough to fix all the coefficients in the homogeneous solution. Only after this is done, we take the single folded limit $u_2\to 1$ to get the result for the three-point function. The reason that we work with the four-point function for the homogeneous solution is that the single folded limit is singular for each term in the homogeneous solution, and the regular three-point limit is achieved only when we take the ``correct'' linear combination of all these terms. 

\paragraph{Particular solution.}As said, we solve the inhomogeneous equation (\ref{eq_BTEs2}) in the single folded limit $u_2=1$. Also writing $u_1=u$, the equation becomes: 
\bge
\label{eq_DupqI}
\mathcal D_{u}^{p_1} \wt{\mathcal I}_{\pm\pm,\text{Inh}}^{p_1p_2}(u,1) = e^{\mp \ii p_{12}\pi/2}\Gamma(5+p_{12})\Big(\FR{u}{2}\Big)^{5+p_{12}}.
\ede
Note that $\wt{\mathcal{I}}_{++,\text{Inh}}^{p_1p_2}$ and $\wt{\mathcal{I}}_{--,\text{Inh}}^{p_1p_2}$ satisfy the equation with the same differential operator on the left-hand side but with the source term differed by a phase. 

We use the following Taylor series as an ansatz for the solution:
\bge
\label{eq_ScalarAnsatz}
\wt{\mathcal I}_{\pm\pm,\text{Inh}}^{p_1p_2}(u,1) = 2^{-5-p_{12}}e^{\mp \ii p_{12}\pi/2}\Gamma(5+p_{12}) \sum_{n=0}^\infty \mathcal{X}_{n} u^{5+p_{12}+n}.
\ede
Then, the equation (\ref{eq_DupqI}) gives rise to the following recursion relation for the Taylor coefficients $\mathcal{X}_n$:
\bge
\mathcal{X}_0 = \FR{1}{(\fr52+p_2)^2+\wt\nu^2},
\qquad
\mathcal{X}_{n+1} = \FR{(n+3+p_2)(n+5+p_{12})}{(n+\fr72+p_2)^2+\wt\nu^2} \mathcal{X}_n.
\ede
Substituting this back into the ansatz (\ref{eq_ScalarAnsatz}), we see that the summation is a standard (generalized) hypergeometric series, which can be directly finished, and we get the following result for the particular solution: 
\bge
\mathcal{X}_n = \FR{(3+p_2)_n(5+p_{12})_n}{(p_2+\fr52-\ii\wt\nu)_{n+1}(p_2+\fr52+\ii\wt\nu)_{n+1}},
\ede
where $(a)_n\equiv\Gamma(a+n)/\Gamma(a)$ is the Pochhammer symbol. Thus 
\begin{align}
\label{eq_IBGs}
\wt{\mathcal I}_{\pm\pm,\text{Inh}}^{p_1p_2}(u,1)=
\FR{e^{\mp \ii p_{12}\pi/2}\Gamma(5+p_{12})u^{5+p_{12}} }{2^{5+p_{12}}[(\fr52+p_2)^2+\wt\nu^2]}  
{}_3\mathrm F_2
\left[\bgm 1,3+p_2,5+p_{12}\\\fr72+p_2-\ii\wt\nu,\fr72+p_2+\ii\wt\nu\edm\middle|u\right].
\end{align} 
This completes our derivation for the particular solution to the bootstrap equation.

\paragraph{Homogeneous solutions.} As mentioned above, the problem of finding the homogeneous solution boils down to a determination of coefficients $\al_{\pm\mp|\aa\bb}$ and $\al_{\pm\pm|\aa\bb}$, as given in (\ref{eq_IpmmpPar}) and (\ref{eq_IpmpmPar}).

We impose the boundary condition in the hierarchical squeezed limit $u_1\ll u_2\ll 1$. In this limit, the tree-seed integral can be directly done. We put this calculation in App.\ \ref{app_hankel}, and here we quote the result: 
\begin{align}
\label{eq_ItildePMSq}
\lim_{u_1\ll u_2\ll 1}\wt{\mathcal I}^{p_1p_2}_{\pm\mp}(u_1,u_2) 
=&~\FR{e^{\mp\ii\bar{p}_{12}\pi/2}}{4\pi}\Big[\wt{\mathcal Y}_+^{p_1}(u_1)+\wt{\mathcal Y}_-^{p_1}(u_1)\Big]\Big[\wt{\mathcal Y}_+^{p_2}(u_2)+\wt{\mathcal Y}_-^{p_2}(u_2)\Big],\\
\label{eq_ItildePPSq}
\lim_{u_1\ll u_2\ll 1}\wt{\mathcal I}^{p_1p_2}_{\pm\pm}(u_1,u_2) 
=&~\FR{\pm\ii e^{\mp\ii p_{12}\pi/2}}{4\pi}\Big[e^{\pi\wt\nu}\wt{\mathcal Y}_\pm^{p_1}(u_1)+e^{-\pi\wt\nu}\wt{\mathcal Y}_{\mp}^{p_1}(u_1)\Big]\Big[\wt{\mathcal Y}_+^{p_2}(u_2)+\wt{\mathcal Y}_-^{p_2}(u_2)\Big].
\end{align}
where $\wt{\mathcal Y}_\pm^p(u)$ is the two independent solutions $\mathcal Y_\pm^p(u)$ given in (\ref{eq_calYpm}) in the squeezed limit $u\to 0$:
\begin{align}
 \wt{\mathcal Y}_\pm^p(u)= 2^{\mp \ii\wt\nu} \Big(\FR{u}{2}\Big)^{5/2+p\pm\ii\wt\nu}\Gamma\Big[\FR52+p+\ii\wt\nu,-\ii\wt\nu\Big].
\end{align}
Given the form of the squeezed-limit solution, we do not have to write down all the coefficients $\al_{\mathsf{cd}|\aa\bb}$ explicitly. Rather, we only need to make the replacement $\wt{\mathcal Y}_\pm^p(u)\to \mathcal Y_\pm^p(u)$ in (\ref{eq_ItildePMSq}) and (\ref{eq_ItildePPSq}), and the results are guaranteed to be the correct homogeneous solutions:
\begin{align}
\label{eq_ItildePM}
\wt{\mathcal I}^{p_1p_2}_{\pm\mp}(u_1,u_2) 
=&~\FR{e^{\mp\ii\bar{p}_{12}\pi/2}}{4\pi}\Big[{\mathcal Y}_+^{p_1}(u_1)+{\mathcal Y}_-^{p_1}(u_1)\Big]\Big[{\mathcal Y}_+^{p_2}(u_2)+{\mathcal Y}_-^{p_2}(u_2)\Big],\\
\label{eq_ItildePP}
\wt{\mathcal I}^{p_1p_2}_{\pm\pm}(u_1,u_2) 
=&~\FR{\pm\ii e^{\mp\ii p_{12}\pi/2}}{4\pi}\Big[e^{\pi\wt\nu}{\mathcal Y}_\pm^{p_1}(u_1)+e^{-\pi\wt\nu}{\mathcal Y}_{\mp}^{p_1}(u_1)\Big]\Big[{\mathcal Y}_+^{p_2}(u_2)+{\mathcal Y}_-^{p_2}(u_2)\Big].
\end{align}

At this point, an important observation is that the squeezed-limit results (\ref{eq_ItildePMSq}) and (\ref{eq_ItildePPSq}) scale with the momentum ratio $u_1$ as $u_{1}^{5/2+p_1\pm \ii\wt\nu}$, which is identical to the scaling of the general homogeneous solution (\ref{eq_calYpm}). On the contrary, the particular solution (\ref{eq_IBGs}) scales as $u_1^{5+p_{12}}$ in the squeezed limit $u_1\ll 1$. [Note that $u_1=u$ in (\ref{eq_IBGs}).] Therefore, for positive real $\wt\nu$, as long as $p_2>-5/2$, the particular solution (\ref{eq_IBGs}) is always subdominant, and the general solutions in (\ref{eq_IpmmpPar}) and (\ref{eq_IpmpmPar}) can be matched to the squeezed-limit result (\ref{eq_ItildePMSq}) and (\ref{eq_ItildePPSq}), respectively. On the other hand, when $p_2\leq -5/2$, the particular solution will dominate the squeezed limit, and there is no way to match the squeezed-limit result. In fact, the Hankel seed integral $\mathcal{I}^{p_1p_2}$ is divergent when $p_{1,2}\leq -5/2$, and the divergence does not cancel out among different SK branches. Physically, the powers $p_{1,2}\leq -5/2$ appear because of too nonlocal interactions, due to which the perturbation theory actually breaks down. Therefore, we shall always assume $p_{1,2}>-5/2$ in this work when $\wt\nu$ is positive real. 

\subsection{Folded limit}

\paragraph{Single folded limit.} Now we consider the single folded limit $u_2\to1$. The general solution (\ref{eq_calYpm}) can become singular in this limit through the hypergeometric factor ${}_2\text{F}_1$, and the singular terms must cancel themselves in (\ref{eq_ItildePM}) and (\ref{eq_ItildePP}), so that only the finite part of the hypergeometric function contributes to the folded limit. Therefore, we only keep the finite part of ${}_2\text{F}_1$ function. In the folded limit $u\to 1$,
\bge
\label{eq_Fin2F1}
\text{Fin}\bigg\{\lim_{u\to 1}{}_2\mathrm F_1 \left[\bgm a,b\\c\edm \middle| u\right]\bigg\} =
\Gamma\left[\bgm c,s\\ a+s,b+s\edm\right],~~~~~~(s\equiv c-a-b\notin\mathbb{Z})
\ede
Here $\text{Fin}\{\cdots\}$ means the finite part of the limit. As indicated, this formula holds only when the balance $s=c-a-b$ is not an integer. The case of $s\in\mathbb{Z}$ can be dealt with by taking the limit. Then, the finite part of the general solution $\mathcal Y_\pm^p(u)$ in (\ref{eq_calYpm}) as $u \to 1$ is:
\bge
\text{Fin}\Big\{\lim_{u\to1}\mathcal Y_{\pm}^p(u)\Big\} = \pm\ii 2^{-5/2-p}\pi^{1/2}
\text{csch}(\pi\wt\nu)\Gamma\left[\bgm-2-p,\fr52+p\pm\ii\wt\nu\\-\fr32-p\pm\ii\wt\nu\edm\right],
\ede
Now, all the general solutions $\mathcal{Y}_\pm^{p_2}(u_2)$ in (\ref{eq_ItildePM}) and (\ref{eq_ItildePP}) come in the combination $\mathcal Y_+^{p_2}(u_2) + \mathcal Y_-^{p_2}(u_2)$. It can be shown that the divergences in the limit $u_2\to 1$ cancel out in this combination, and we have the following regular limit:
\bge
\label{eq_uPanduM}
\lim_{u\to1}\Big[\mathcal Y_+^{p}(u) + \mathcal Y_-^{p}(u)\Big]  =2^{-3/2-p}\pi^{1/2}\Gamma\left[\bgm \fr52+p-\ii\wt\nu,\fr52+p+\ii\wt\nu\\
3+p\edm\right].
\ede
Substituting this result back into (\ref{eq_ItildePM}) and (\ref{eq_ItildePP}), we find the correct homogeneous solution in the single folded limit. The explicit results are summarized  in (\ref{eq_IPPresult}) and (\ref{eq_IPMresult}) at the end of this section.

\paragraph{Double-folded limit.}
Now we consider the double folded limit, in order to get an expression for the two-point correlator. The double folded limit is obtained by further taking $u\to 1$ in $\wt{\mathcal{I}}_{\aa\bb}^{p_1p_2}(u,1)$. Remarkably, all of the four seed integral $\wt{\mathcal I}_{\aa\bb}^{p_1p_2}(u,1)$ are individually regular in the limit of $u\to 1$. First, the regularity of $\wt{\mathcal I}_{\pm\mp}^{p_1p_2}(u,1)$ is again due to the combination ${\mathcal Y}_+^{p_1}(u)+{\mathcal Y}_-^{p_1}(u)$, which we have shown to be regular in the limit of $u\to 1$ in the previous subsection. In fact, by direct substitution of (\ref{eq_uPanduM}) in $\wt{\mathcal I}_{\pm\mp}^{p_1p_2}(u,1)$ in  (\ref{eq_IPMresult}), we immediately get:
\begin{align}
   \wt{\mathcal I}_{\pm\mp}^{p_1p_2}(1,1)
  =&~\FR{e^{\mp\ii\bar{p}_{12}\pi/2}}{2^{5+p_{12}} }\Gamma\left[\bgm \fr52+p_1-\ii\wt\nu,\fr52+p_1+\ii\wt\nu,\fr52+p_2-\ii\wt\nu,\fr52+p_2+\ii\wt\nu\\
3+p_1,3+p_2\edm\right].
\end{align}
On the other hand, in $\wt{\mathcal I}_{\pm\pm}^{p_1p_2}(u,1)$ in (\ref{eq_IPPresult}), the combination $e^{\pi\wt\nu}{\mathcal Y}_+^{p_1}(u)+e^{-\pi\wt\nu}{\mathcal Y}_-^{p_1}(u)$ is divergent when $u\to 1$, and this divergence is to be canceled by another divergent piece from the ${}_3\text{F}_2$ function in (\ref{eq_IPPresult}), as a consequence of the Bunch-Davies initial condition for all the mode functions. Therefore, we shall only retain the finite part of each of these two pieces, and add them together to get the result in the double folded limit.

The finite part of the combination $e^{\pi\wt\nu}{\mathcal Y}_+^{p_1}(u)+e^{-\pi\wt\nu}{\mathcal Y}_-^{p_1}(u)$ can again be obtained by using the formula in (\ref{eq_Fin2F1}). To obtain the finite part of the ${}_3\text{F}_2$ function, we use the following formula \cite{doi:10.1137/0518089}:
\bge
\label{eq_3F2finite}
\text{Fin}\bigg\{\lim_{u\to 1}{}_3\mathrm F_2 \left[\bgm a,b,c\\d,e\edm \middle| u\right]\bigg\} =
\Gamma\left[\bgm d,e,s\\ c,a+s,b+s\edm\right]
{}_3\mathrm F_2 \left[\bgm d-c,e-c,s\\a+s,b+s\edm \middle| 1\right] ,
\ede
where $s\equiv d+e-a-b-c$ is the balance of the hypergeometric function. Once again, the above formula holds when $s$ is not an integer, and the integer case can be dealt with by taking the limit. Therefore, for the ${}_3\text{F}_2$ function in the inhomogeneous solution (\ref{eq_IBGs}), we can make the following assignment of parameters,
\bge
a=5+p_{12},\quad b=3+p_2,\quad c=1,
\quad d=\FR72+p_2-\ii\wt\nu,\quad e=\FR72+p_2+\ii\wt\nu,\quad
s=-2-p_1.
\ede
Combined with the finite part of the homogeneous solution as shown in (\ref{eq_IpmpmPar}), we obtain the double-folded limit of the $\wt{\mathcal I}_{\pm\pm}^{p_1p_2}$ as follows:
\begin{align}
\label{eq_I2ptPP2ptSimp}
  &\wt{\mathcal I}_{\pm\pm}^{p_1p_2}(1,1)\n\\
  =&~\FR{\pm\ii e^{\mp\ii {p}_{12}\pi/2}[1\mp\ii\cot(\pi p_1)]\cosh\pi\wt\nu}{2^{5+p_{12}}}\Gamma\left[\bgm\fr52+p_1-\ii\wt\nu,\fr52+p_1+\ii\wt\nu, \fr52+p_2-\ii\wt\nu,\fr52+p_2+\ii\wt\nu\\
3+p_1,3+p_2\edm\right] \n\\
   &+\FR{e^{\mp \ii p_{12}\pi/2}\Gamma(5+p_{12})}{2^{5+p_{12}}}  
 {}_3\mathcal{F}_2
\left[\bgm  \fr52+p_2-\ii\wt\nu,\fr52+p_2+\ii\wt\nu,-2-p_1\\ 3+p_2,1-\bar{p}_{12}\edm\middle|1\right],
\end{align}

There are a large number of relations among hypergeometric functions of argument unity. With these relations we can recast the result into a somewhat simpler form. For example, we can use the following relation (Eq.\ 4.3.4.4 of \cite{Slater:1966}):
\begin{align}
\label{eq_3F2transf}
&~{}_3\mathrm F_2\left[\bgm a,b,d+e-a-b-1\\ d,e \edm\middle|1\right]
=\Gamma\left[\bgm
d,e,d-a-b,e-a-b\\d-a,d-b,e-a,e-b
\edm\right]\n\\
&+\FR{1}{a+b-d}\Gamma\left[\bgm
d,e\\a,b,d+e-a-b\edm\right]
{}_3\mathrm F_2\left[\bgm d-a,d-b,1\\1+d-a-b,d+e-a-b\edm\middle|1\right].
\end{align}
Then, with the following assignment of parameters:
\bge
a=\FR52+p_2-\ii\wt\nu,\quad
b=-2-p_1,\quad
d=3+p_2,\quad
e=1-p_1+p_2,
\ede
we obtain: 
\begin{align}
  &\wt{\mathcal I}_{\pm\pm}^{p_1p_2}(1,1)
  =\FR{\pm\ii e^{\mp\ii {p}_{12}\pi/2}e^{-\pi\wt\nu}}{2^{5+p_{12}}}\Gamma\left[\bgm\fr52+p_1-\ii\wt\nu,\fr52+p_1+\ii\wt\nu, \fr52+p_2-\ii\wt\nu,\fr52+p_2+\ii\wt\nu\\
3+p_1,3+p_2\edm\right] \n\\
   &-\FR{e^{\mp \ii p_{12}\pi/2}}{2^{5+p_{12}}}\Gamma\Big[5+p_{12},\fr52+p_1\pm\ii\wt\nu,\fr52+p_2\pm\ii\wt\nu\Big]  
 {}_3\wt{\mathrm{F}}_2
\left[\bgm  5+p_{12},\fr12\pm\ii\wt\nu,1\\ \fr72+p_1\pm\ii\wt\nu,\fr72+p_2\pm\ii\wt\nu\edm\middle|1\right].
\end{align}
This result has the advantage that it is manifestly symmetric in $p_1$ and $p_2$, and is regular when $p_1$ and $p_2$ take generic integer values. 

\subsection{Summary}
\label{sec_hankel_summary}
 
At this point, we have finished the computation of the Hankel seed integral $\wt{\mathcal I}_{\aa\bb}^{p_1p_2}(u,1)$ in the single folded limit. We present the full result below.
\begin{keyeqn}
\begin{align}
\label{eq_IPPresult}
  \wt{\mathcal I}_{\pm\pm}^{p_1p_2}(u,1)
  =&~\FR{\pm\ii e^{\mp\ii {p}_{12}\pi/2}}{2^{7/2+p_2}\pi^{1/2}}\Gamma\left[\bgm \fr52+p_2-\ii\wt\nu,\fr52+p_2+\ii\wt\nu\\
3+p_2\edm\right]\Big[e^{\pi\wt\nu}{\mathcal Y}_\pm^{p_1}(u)+e^{-\pi\wt\nu}{\mathcal Y}_\mp^{p_1}(u)\Big]\n\\
   &+\FR{e^{\mp \ii p_{12}\pi/2}\Gamma(5+p_{12})u^{5+p_{12}} }{2^{5+p_{12}}[(\fr52+p_2)^2+\wt\nu^2]}  
{}_3\mathrm F_2
\left[\bgm 1,3+p_2,5+p_{12}\\\fr72+p_2-\ii\wt\nu,\fr72+p_2+\ii\wt\nu\edm\middle|u\right],\\
\label{eq_IPMresult}
  \wt{\mathcal I}_{\pm\mp}^{p_1p_2}(u,1)
  =&~\FR{e^{\mp\ii\bar{p}_{12}\pi/2}}{2^{7/2+p_2}\pi^{1/2}}\Gamma\left[\bgm \fr52+p_2-\ii\wt\nu,\fr52+p_2+\ii\wt\nu\\
3+p_2\edm\right]\Big[{\mathcal Y}_+^{p_1}(u)+{\mathcal Y}_-^{p_1}(u)\Big] ,
\end{align}
\end{keyeqn}
where $\mathcal Y_\pm^{p}(u)$ is a pair of independent solutions to the homogeneous bootstrap equation:
\bge 
\mathcal Y_\pm^{p}(u) = 2^{\mp\ii\wt\nu}\Big(\FR{u}2\Big)^{5/2+p\pm\ii\wt\nu}
\Gamma\Big[\FR52+p\pm\ii\wt\nu,\mp\ii\wt\nu\Big]
{}_2\mathrm F_1\left[\bgm\fr52+p\pm\ii\wt\nu,\fr12\pm\ii\wt\nu\\1\pm2\ii\wt\nu\edm\middle|u\right].
\ede
With these results, one can easily generate explicit analytical expressions for a large class of three-point correlators, as discussed in Sec.\ \ref{sec_reduce}.

 The result for the two-point function can be read from the Hankel seed integral in the double folded limit. The results are:
\begin{keyeqn}
\begin{align}
\label{eq_H2ptResultPP}
  &\wt{\mathcal I}_{\pm\pm}^{p_1p_2}(1,1)
  =\FR{\pm\ii e^{\mp\ii {p}_{12}\pi/2}e^{-\pi\wt\nu}}{2^{5+p_{12}}}\Gamma\left[\bgm\fr52+p_1-\ii\wt\nu,\fr52+p_1+\ii\wt\nu, \fr52+p_2-\ii\wt\nu,\fr52+p_2+\ii\wt\nu\\
3+p_1,3+p_2\edm\right] \n\\
   &-\FR{e^{\mp \ii p_{12}\pi/2}}{2^{5+p_{12}}}\Gamma\Big[5+p_{12},\fr52+p_1\pm\ii\wt\nu,\fr52+p_2\pm\ii\wt\nu\Big]  
  {}_3\wt{\mathrm{F}}_2 
\left[\bgm  5+p_{12},\fr12\pm\ii\wt\nu,1\\ \fr72+p_1\pm\ii\wt\nu,\fr72+p_2\pm\ii\wt\nu\edm\middle|1\right],\\
\label{eq_H2ptResultPM}
 &   \wt{\mathcal I}_{\pm\mp}^{p_1p_2}(1,1)
  =\FR{e^{\mp\ii\bar{p}_{12}\pi/2}}{2^{5+p_{12}} }\Gamma\left[\bgm \fr52+p_1-\ii\wt\nu,\fr52+p_1+\ii\wt\nu,\fr52+p_2-\ii\wt\nu,\fr52+p_2+\ii\wt\nu\\
3+p_1,3+p_2\edm\right].
\end{align}
\end{keyeqn}

\section{Bootstrapping Whittaker Seed Integrals} 
\label{sec_whittaker}

Now we consider the case where the intermediate massive particle has a boost-breaking chemical potential. This scenario is particularly interesting for CC phenomenology due to its exponentially enhanced signal and also its parity-violating nature. Mathematically, the mode functions for such fields involve the Whittaker W function, which needs a separate treatment from the previously considered Hankel type. In this section, we shall consider the Whittaker case and go through the same procedure as the last section. We shall be briefer in this section since most of the intermediate steps are in parallel with the previous section. 

\subsection{Whittaker seed integral and its bootstrap equation}

As introduced in Sec.\ \ref{sec_reduce}, the Whittaker seed integral (\ref{eq_SeedIntW}) is defined based on the SK integral for the tree-level four-point correlator with the exchange of a spin-1 particle of mass parameter $\wt\nu$ and helicity-dependent chemical potential $\wt\mu$. In parallel with the previous section, we first derive the bootstrap equations for this seed integral with $r$-variables, and then change to $u$-variables. Thus we rewrite the Whittaker seed integral (\ref{eq_SeedIntW}) in the following way:
\begin{align}
\label{eq_seedIntWr}
\mathcal{I}^{(h)p_1p_2}_{\aa\bb}(r_1,r_2) 
= -\aa\bb\,r_1^{1+p_1}r_2^{1+p_2}\int_0^\infty \di z_1\di z_2\,
z_1^{p_1}z_2^{p_2} e^{-\ii\aa z_1-\ii\bb z_2}\wh D_{\aa\bb}^{(h)}(r_1z_1,r_2z_2).
\end{align}
In the seed integral above, $\wh D_{\aa\bb}^{(h)}(z_1,z_2)$ is again a hatted propagator, this time built from the helicity-$h$ component ($h=\pm 1$) of the massive spin-1 propagator $D_{\aa\bb}^{(h)}(k;\tau_1,\tau_2)$:
\bge
  \wh D_{\aa\bb}^{(h)}(z_1,z_2)=k D_{\aa\bb}^{(h)}(k;\tau_1,\tau_2).
\ede
Again, we use the dimensionless and positive variables $z_i=-k\tau_i$ $(i=1,2)$. Here the four Schwinger-Keldysh propagators $D_{\aa\bb}^{(h)}(k;\tau_1,\tau_2)$ are again related to the Wightman functions $D_>^{(h)}$ and $D_<^{(h)}=D_>^{(h)*}$, and the ``greater'' Wightman function is given by
\begin{align}
  D^{(h)}_>(k;\tau_1,\tau_2)=&~ \FR{e^{-\pi h\wt\mu}}{2k}\mathrm{W}_{\ii h\wt\mu,\ii\wt\nu}(2\ii k\tau_1)\mathrm{W}_{-\ii h\wt\mu,\ii\wt\nu}(-2\ii k\tau_2).
\end{align}
Let us emphasize that, although we are building the Whittaker seed integral from the spin-1 propagators, this seed integral can be used to compute correlators involving higher spin states. Very often, the boost-breaking chemical potential changes the dispersion of a spin-$s$ field in such a way that its highest or lowest helicity component ($h=\pm s$) is most enhanced. For such states, the mode function is identical to the $h=\pm 1$ states of the spin-1 field up to a prefactor, and thus above seed integral is perfectly applicable to these cases. One only needs to be careful that the mass parameter $\wt\nu$ is related to the mass $m$ of a spin-$s$ state via $\wt\nu=\sqrt{m^2-(s-1/2)^2}$.

The equations of motion satisfied by the propagator $D_{\aa\bb}^{(h)}(k;\tau_1,\tau_2)$ then lead to the following set of equations satisfied by the hatted propagators:
\begin{align}
&(r_1^2\pd_{r_1}^2+r_1^2z_1^2 +2h\wt\mu r_1z_1+m^2)\wh D_{\pm\mp}^{(h)}(r_1z_1,r_2z_2) = 0,\\
&(r_1^2\pd_{r_1}^2+r_1^2z_1^2 +2h\wt\mu r_1z_1+m^2)\wh D_{\pm\pm}^{(h)}(r_1z_1,r_2z_2) = \mp \ii r_1r_2z_1z_2\de(r_1z_1-r_2z_2).
\end{align} 
Therefore, we insert the differential operator $r_1^2\pd_{r_1}^2+r_1^2z_1^2 +2h\wt\mu r_1z_1+m^2$ in front of the hatted propagator in the Whittaker seed integral (\ref{eq_seedIntWr}), to derive the bootstrap equation. On one hand, the differential operator reduces the integral to either 0 or a local term, which becomes the right-hand side of the bootstrap equation. On the other hand, we commute the differential operator with the integrals over $z_{1,2}$ with the help of (\ref{eq_zIBP}) and (\ref{eq_z2IBP}), and end up with a new differential operator acting on the whole seed integral, which is the left-hand side of the bootstrap equation. The result is:
\begin{align}
&\mathcal{D}_{\pm,r_1}^{p_1} \mathcal I_{\pm\mp}^{(h)}(r_1,r_2)= 0,\\
&\mathcal{D}_{\pm,r_1}^{p_1}  \mathcal I_{\pm\pm}^{(h)}(r_1,r_2)
=-e^{\mp\ii p_{12}\pi/2}\Gamma(3+p_{12})\Big(\FR{r_1r_2}{r_1+r_2}\Big)^{3+p_{12}};\\
&\mathcal{D}_{\pm,r}^p\equiv (r^2-r^4)\pd_{r}^2-\Big[2(1+p)r\pm 2\ii h\wt\mu r^2+2r^3\Big]\pd_{r}
+ \Big[\wt\nu^2+\fr{(3+2p)^2}4\Big]. 
\end{align}
Next we use the new variables $u_i = (2r_i)/(1+r_i)$ in place of $r_i$ $(i=1,2)$, in terms of which the seed integral is rewritten as 
\bge
\wt{\mathcal I}_{\aa\bb}^{(h)p_1p_2}(u_1,u_2)\equiv \mathcal I_{\aa\bb}^{(h)p_1p_2}\big(r_1(u_1),r_2(u_2)\big)
\ede
Then we can derive the following set of equations with respect to the $u$-variable:
\begin{align}
\label{eq_BTEV1}
&\mathcal D_{\pm,u_1}^{p_1}\wt{\mathcal I}_{\pm\mp}^{(h)}(u_1,u_2)= 0,\\
\label{eq_BTEV2}
&\mathcal D_{\pm,u_1}^{p_1}\wt{\mathcal I}_{\pm\pm}^{(h)}(u_1,u_2)
=-e^{\mp\ii p_{12}\pi/2}\Gamma(3+p_{12})\Big(\FR{u_1u_2}{2(u_1+u_2-u_1u_2)}\Big)^{3+p_{12}};\\
\label{eq_Dpmu}
&\mathcal D_{\pm,u}^p = (u^2-u^3)\pd_u^2 - \Big[(2+2p)u+(\pm \ii h\wt\mu-p)u^2\Big]\pd_u + \Big[\wt\nu^2+\big(p+\fr32\big)^2\Big].
\end{align}
These are the bootstrap equations for the Whittaker seed integral, from which we shall bootstrap the three-point and two-point functions in the following subsections. 

\subsection{Solving the bootstrap equation}

In parallel with the procedure in the last section, we now solve the bootstrap equations (\ref{eq_BTEV1}) and (\ref{eq_BTEV2}). Again, the solution to the homogeneous equation (\ref{eq_BTEV1}) can be written as a linear combination of the two independent solutions
\begin{align}
\label{eq_IhPMpar}
  \wt{\mathcal I}_{\pm\mp}^{(h)p_1p_2}(u_1,u_2)=\sum_{\aa,\bb=\pm}\be_{\pm\mp|\aa\bb}\,\mathcal{U}_{\pm|\aa}^{p_1}(u_1)\,\mathcal{U}_{\mp|\bb}^{p_2}(u_2),
\end{align}
with $\mathcal{U}^{p}_{\aa|\bb}(u)$ ($\bb=\pm1$) the two independent solutions to the following equations:
\begin{align}
  \mathcal{D}_{\aa,u}^p\mathcal{U}^p_{\aa|\bb}(u)=0.
\end{align}
Here $\mathcal{D}_{\aa,u}^p$ are the differential operators in (\ref{eq_Dpmu}). Note that we have two distinct operators $\mathcal{D}_{\aa,u}^p$, labeled by $\aa=\pm$. For each fixed choice of $\aa=\pm$, there are a pair of independent solutions, labeled by $\bb=\pm$, hence the notation $\mathcal{U}^p_{\aa|\bb}$ for the solutions. Explicitly, we can choose $\mathcal{U}^p_{\aa|\bb}$ to be:
\bge
\mathcal U_{\aa|\bb}^p(u) = \ii\,\aa\bb\, 2^{\ii\aa\bb\wt\nu} \pi \text{csch}(2\pi\wt\nu)  \Big(\FR u2\Big)^{3/2+p+ \ii\aa\bb\wt\nu} {}_2\mathcal F_1\left[\bgm \fr32+p+\ii\aa\bb\wt\nu,\fr12+\ii\aa h\wt\mu+\ii\aa\bb\wt\nu\\1+2\ii\aa\bb\wt\nu \edm\middle| u \right].
\ede
Here, again, we have included some numerical factors for later convenience. The coefficients $\be_{\pm\mp|\aa\bb}$ in (\ref{eq_IhPMpar}) should be determined by the boundary conditions in the squeezed limit, as will be detailed below.

For the inhomogeneous equation (\ref{eq_BTEV2}), the solution is a sum of a particular solution $\wt{\mathcal I}_{\pm\pm,\text{Inh}}^{(h)p_1p_2}(u_1,u_2)$ and a homogeneous solution which is again a linear combination of independent solutions:
\begin{align}
  \wt{\mathcal I}_{\pm\pm}^{(h)p_1p_2}(u_1,u_2)=\wt{\mathcal I}_{\pm\pm,\text{Inh}}^{(h)p_1p_2}(u_1,u_2)+\sum_{\aa,\bb=\pm}\be_{\pm\pm|\aa\bb}\,\mathcal{U}^{p_1}_{\pm|\aa}(u_1)\,\mathcal{U}^{p_2}_{\pm|\bb}(u_2).
\end{align}
Below we will first find the particular solution $\wt{\mathcal I}_{\pm\pm,\text{Inh}}^{(h)p_1p_2}(u_1,u_2)$, and then determine the homogeneous solutions by matching the results in the squeezed limit.

\paragraph{Particular solution.} 
Similar to the previous section, we do not pursue the most general result for the particular solution. Rather, we shall set $u_2=1$ and write $u_1=u$ in the inhomogeneous equation (\ref{eq_BTEV2}):
\begin{align}
&\mathcal D_{\pm,u}^{p_1} \wt{\mathcal I}^{(h)p_1p_2}_{\pm\pm,\text{Inh}}(u,1) = -e^{\mp\ii p_{12}\pi/2}\Gamma(3+p_{12})\Big(\FR u2\Big)^{3+p_{12}}.
\end{align}
This equation can again be solved by trying the following series ansatz:
\bge
\wt{\mathcal I}^{(h)p_1p_2}_{\pm\pm,\text{Inh}}(u,1) = -2^{-3-p_{12}}e^{\mp\ii p_{12}\pi/2}\Gamma(3+p_{12})
 \sum_{n=0}^\infty \mathcal{V}_{n,\pm} u^{3+p_{12}+n}.
\ede
The differential equation then leads to the following recursion relations for the series coefficients:
\bge
\mathcal{V}_{0,\pm} = \FR{1}{(\fr32+p_2)^2+\wt\nu^2},
\qquad
\mathcal{V}_{n+1,\pm} = \FR{(n+2+p_2\pm\ii h\wt\mu)(n+3+p_{12})}{(n+\fr52+p_2)^2+\wt\nu^2} \mathcal{V}_{n,\pm},
\ede
and the recursion relation can be directly solved to get a general term formula: 
\bge
\mathcal{V}_{n,\pm} = \FR{(2+p_2\pm\ii h\wt\mu)_n(3+p_{12})_n}{(p_2+\fr32-\ii\wt\nu)_{n+1}(p_2+\fr32+\ii\wt\nu)_{n+1}}.
\ede
This shows that the series ansatz is again a standard hypergeometric series. Finishing the summation, we get:
\bge
\wt{\mathcal I}^{(h)p_1p_2}_{\pm\pm,\text{Inh}}(u,1)=
-  
\FR{e^{\mp\ii p_{12}\pi/2}\Gamma(3+p_{12})u^{3+p_{12}}}{2^{3+p_{12}}[(\fr32+p_2)^2+\wt\nu^2]}
{}_3\mathrm F_2
\left[\bgm 1,2+p_2\pm\ii h\wt\mu,3+p_{12}\\\fr52+p_2-\ii\wt\nu,\fr52+p_2+\ii\wt\nu\edm\middle|u\right].
\ede
This completes our computation of the particular solution. 
\paragraph{Homogeneous solutions.} To determine the homogeneous solutions in both $\wt{\mathcal{I}}_{\pm\mp}^{(h)p_1p_2}(u_1,u_2)$ and $\wt{\mathcal{I}}_{\pm\pm}^{(h)p_1p_2}(u_1,u_2)$, we again compute these integrals in the squeezed limit $u_1\ll u_2\ll 1$. Some details are given in App.\ \ref{app_whittaker}, and the results are shown in (\ref{eq_IhSqLimit}). By matching the results in the squeezed limit, we can determine the coefficients $\be_{\aa\bb|\cc\dd}$:
\begin{align}
\label{eq_betaCoef1}
  &\be_{\pm\mp|++}=\be_{\pm\mp|--}=\be_{\pm\mp|+-}=\be_{\pm\mp|-+}=\FR{e^{\mp\ii \bar p_{12}\pi/2}e^{-\pi h\wt\mu}}{2\pi^2}(\cosh 2\pi\wt\mu + \cosh 2\pi\wt\nu),\\
\label{eq_betaCoef2}
  &\be_{\pm\pm|++}=\be_{\pm\pm|+-}=\FR{\mp\ii e^{\mp\ii p_{12}\pi/2}e^{-\pi (h\wt\mu-\wt\nu)}\cosh[\pi(h\wt\mu+\wt\nu)]}{\pi \Gamma\big[\fr12\pm\ii h\wt\mu-\ii\wt\nu,\fr12\pm\ii h\wt\mu+\ii\wt\nu\big]},\\
\label{eq_betaCoef3}
  &\be_{\pm\pm|-+}=\be_{\pm\pm|--}=\FR{\mp\ii e^{\mp\ii p_{12}\pi/2}e^{-\pi (h\wt\mu+\wt\nu)}\cosh[\pi(h\wt\mu-\wt\nu)]}{\pi \Gamma\big[\fr12\pm\ii h\wt\mu-\ii\wt\nu,\fr12\pm\ii h\wt\mu+\ii\wt\nu\big]}.
\end{align}

 
\subsection{Folded limit}

\paragraph{Single-folded limit.} Now we consider the folded limits of the Whittaker seed integrals. First, to obtain the three-point function, we take the single-folded limit $u_2\to 1$. As in the previous section, each of the solutions $\mathcal U_{\aa|\bb}^p(u)$ is singular when $u\to 1$. However, the singular terms must cancel out in the final expression, as a result of choosing the Bunch-Davies initial condition. Therefore, we only need to retain the finite parts of the solutions $\mathcal U_{\aa|\bb}^p(u)$ in the folded limit $u\to 1$. For notational simplicity, we define the finite part of $\mathcal U_{\aa|\bb}^p(1)$ as: 
\bge
U_{\aa|\bb}^p \equiv\text{Fin}\Big\{\mathcal U_{\aa|\bb}^p(1)\Big\} =\FR{\ii\aa\bb\pi}{2^{3/2+p}\sinh(2\pi\wt\nu)}
\Gamma\left[\bgm
\fr32+p+\ii\aa\bb\wt\nu,\fr12+\ii\aa h\wt\mu+\ii\aa\bb\wt\nu,-1-p-\ii\aa h\wt\mu\\-\fr12-p+\ii\aa\bb\wt\nu,\fr12-\ii\aa h\wt\mu+\ii\aa\bb\wt\nu
\edm\right].
\ede  
Looking at the structure of the coefficients $\be_{\aa\bb|\cc\dd}$ in (\ref{eq_betaCoef1}) - (\ref{eq_betaCoef3}), we see that the $u_2$-dependence in the full homogeneous solution always appears in the combination ${\mathcal U}_{\pm|+}^{p_2}(u_2) + {\mathcal U}_{\pm|-}^{p_2}(u_2)$. It can be shown that the singular terms in ${\mathcal U}_{\aa|\bb}^{p_2}(u_2)$ always cancel in this combination, and therefore, we only need the following finite result:
\begin{align}
& U_{\pm|+}^p + U_{\pm|-}^p
 = \FR{\pm\ii\pi\Gamma(-1-p\mp\ii h\wt\mu)}{2^{3/2+p}\sinh(2\pi\wt\nu)} 
 \Gamma\left[\bgm
\fr32+p\pm\ii\wt\nu,\fr12\pm\ii h\wt\mu\pm\ii\wt\nu\\
-\fr12-p\pm\ii\wt\nu,\fr12\mp\ii h\wt\mu\pm\ii\wt\nu
\edm\right]
+(\wt\nu\to-\wt\nu) .
\end{align}
Then, the single-folded limit of the Whittaker seed integral can be written in the following way: 
\begin{align}
\label{eq_Ih3ptU1}
  \wt{\mathcal I}_{\pm\mp}^{(h)p_1p_2}(u,1)=&~\be_{\pm\mp|++}\Big(U_{\mp|+}^{p_2}+U_{\mp|-}^{p_2}\Big)\Big[\,\mathcal{U}_{\pm|+}^{p_1}(u)+\mathcal{U}_{\pm|-}^{p_1}(u)\Big],\\
\label{eq_Ih3ptU2}
  \wt{\mathcal I}_{\pm\pm}^{(h)p_1p_2}(u,1)=&~\Big(U_{\mp|+}^{p_2}+U_{\mp|-}^{p_2}\Big)\Big[\be_{\pm\pm|++}\,\mathcal{U}_{\pm|+}^{p_1}(u)+\be_{\pm\pm|-+}\mathcal{U}_{\pm|-}^{p_1}(u)\Big]+\wt{\mathcal I}^{(h)p_1p_2}_{\pm\pm,\text{Inh}}(u,1).
\end{align}
All quantities in these expressions have been calculated; A simple substitution then gives us the explicit expressions. We present these explicit results at the end of this section.

\paragraph{Double-folded limit.}
Next we consider the double folded limit, which will give us an expression for the two-point function. Formally, this limit can be reached by taking $u\to 1$ in (\ref{eq_Ih3ptU1}) and (\ref{eq_Ih3ptU2}). That is: 
\begin{align}
\label{eq_Ih2ptU1}
  \wt{\mathcal I}_{\pm\mp}^{(h)p_1p_2}(u,1)=&~\be_{\pm\mp|++}\Big(U_{\pm|+}^{p_1}+U_{\pm|-}^{p_1}\Big)\Big(U_{\mp|+}^{p_2}+U_{\mp|-}^{p_2}\Big),\\
\label{eq_Ih2ptU2}
  \wt{\mathcal I}_{\pm\pm}^{(h)p_1p_2}(u,1)=&~\Big(\be_{\pm\pm|++}U_{\pm|+}^{p_1}+\be_{\pm\pm|-+}U_{\pm|-}^{p_1}\Big)\Big(U_{\mp|+}^{p_2}+U_{\mp|-}^{p_2}\Big)+\text{Fin}\Big\{\wt{\mathcal I}^{(h)p_1p_2}_{\pm\pm,\text{Inh}}(1,1)\Big\}.
\end{align}
To get the finite part of the particular solution in the last line, we again use the formula (\ref{eq_3F2finite}) with the following assignment of parameters:
\begin{align}
 a=3+p_{12},\quad b=2+p_2+\ii h\wt\mu,\quad c=1, \quad 
 d=\FR52+p_2-\ii\wt\nu,\quad e=\FR52+p_2+\ii\wt\nu.
\end{align}
Therefore, the finite part of the background is:
\begin{align}
\text{Fin}\Big\{\wt{\mathcal I}^{(h)p_1p_2}_{\pm\pm,\text{Inh}}(1,1)\Big\}
=&-\FR{e^{\mp\ii p_{12}\pi/2}\Gamma(3+p_{12})}{2^{3+p_{12}}}  
 {}_3\mathcal F_2
\left[\bgm \fr32+p_2-\ii\wt\nu,\fr32+p_2+\ii\wt\nu,-1-p_1-\ii h\wt\mu\\2+p_2-\ii h\wt\mu,1-\bar p_{12}\edm\middle|1\right].
\end{align}
This expression can again be put into a better form by using (\ref{eq_3F2transf}) with the following assignment of parameters: 
\bge
a=\FR32+p_2-\ii\wt\nu,\quad
b=-1-p_1-\ii h\wt\mu,\quad
d=2+p_2-\ii h\wt\mu,\quad
e=1-p_1+p_2.
\ede
The resulting expression will be summarized in the next subsection.

\subsection{Summary}

Now we summarize the results of this section by presenting the explicit expressions for the Whittaker seed integral in the folded limits. This include the three-point function $\wt{\mathcal I}_{\aa\bb}^{(h)p_1p_2}(u,1)$ and the two-point function $\wt{\mathcal I}_{\aa\bb}^{(h)p_1p_2}(1,1)$. For the three-point function, we have:
\begin{keyeqn}
\begin{align}
\label{eq_IhPMresult}
\wt{\mathcal I}_{\pm\mp}^{(h)p_1p_2}(u,1) 
=&~\bigg\{\FR{\pm\ii e^{\mp\ii \bar p_{12}\pi/2}e^{-\pi h\wt\mu}}{2^{3/2+p_2}\sinh(2\pi\wt\nu)}
\Gamma\left[\bgm -1-p_2\pm\ii h\wt\mu,\fr32+p_2\pm\ii\wt\nu\\ 
\fr12\pm\ii h\wt\mu-\ii\wt\nu,\fr12\pm\ii h\wt\mu+\ii\wt\nu,-\fr12-p_2\pm\ii\wt\nu\edm\right]\n\\
&~\times\cosh\big[\pi(h\wt\mu+\wt\nu)\big]+(\wt\nu\to-\wt\nu)
\bigg\} \Big[\,\mathcal U_{\pm|+}^{p_1}(u)+\mathcal U_{\pm|-}^{p_1}(u)\Big],\\
\label{eq_IhPPresult}
\wt{\mathcal I}_{\pm\pm}^{p_1p_2}(u,1) 
=&~\bigg\{\FR{e^{\mp\ii p_{12}\pi/2}e^{-\pi h\wt\mu}\cosh[\pi(h\wt\mu-\wt\nu)]}{2^{3/2+p_2}\pi\sinh(2\pi\wt\nu)} 
\Gamma\left[\bgm
-1-p_2\mp\ii h\wt\mu,\fr32+p_2\pm\ii\wt\nu\\
-\fr12-p_2\pm\ii\wt\nu
\edm\right]
+(\wt\nu\to-\wt\nu)
\bigg\}\n\\
&\times \Big\{e^{\pi\wt\nu}\cosh\big[\pi(h\wt\mu+\wt\nu)\big]\mathcal U_{\pm|+}^{p_1}(u_1)+(\wt\nu\to-\wt\nu)\Big\}\n\\
&~-\FR{e^{\mp\ii p_{12}\pi/2}\Gamma(3+p_{12})u^{3+p_{12}}}{2^{3+p_{12}}[(\fr32+p_2)^2+\wt\nu^2]}
{}_3\mathrm F_2
\left[\bgm 1,2+p_2\pm\ii h\wt\mu,3+p_{12}\\\fr52+p_2-\ii\wt\nu,\fr52+p_2+\ii\wt\nu\edm\middle|u\right].
\end{align}
\end{keyeqn}
Here again the momentum ratio $u=2k_3/k_{123}$, and $\mathcal{U}_{\aa|\bb}^p(u)$ are the independent solutions to the homogeneous bootstrap equations, whose explicit expression are:
\bge
\mathcal U_{\aa|\bb}^p(u) = \ii\,\aa\bb\, 2^{\ii\aa\bb\wt\nu} \pi \text{csch}(2\pi\wt\nu)  \Big(\FR u2\Big)^{3/2+p+ \ii\aa\bb\wt\nu} {}_2\mathcal F_1\left[\bgm \fr32+p+\ii\aa\bb\wt\nu,\fr12+\ii\aa h\wt\mu+\ii\aa\bb\wt\nu\\1+2\ii\aa\bb\wt\nu \edm\middle| u \right].
\ede
For the two-point function, we have
\begin{keyeqn}
\begin{align}
\label{eq_Ih2ptResultPM}
\mathcal I_{\pm\mp}^{(h)p_1p_2}(1,1) =&~\FR{e^{\mp\ii \bar p_{12}\pi/2}e^{-\pi h\wt\mu}}{2^{3+p_{12}}}
\Gamma\left[\bgm\fr32+p_1-\ii\wt\nu,\fr32+p_1+\ii\wt\nu,\fr32+p_2-\ii\wt\nu,\fr32+p_2+\ii\wt\nu\\
2+p_1\pm\ii h\wt\mu,2+p_2\mp\ii h\wt\mu\edm\right],\\
\label{eq_Ih2ptResultPP}
\mathcal I_{\pm\pm}^{(h)p_1p_2}(1,1) =&~\FR{\mp\ii e^{\mp\ii p_{12}\pi/2}
}{2^{3+p_{12}}}\Gamma\Big[\FR32+p_1\pm\ii\wt\nu,\FR32+p_2\pm\ii\wt\nu\Big] \n\\
&\times \bigg\{\FR{e^{-2\pi h\wt\mu}+e^{-2\pi\wt\nu}}{2\pi}\Gamma\left[\bgm\fr32+p_1\mp\ii\wt\nu,\fr32+p_2\pm\ii\wt\nu,\fr12\pm\ii h\wt\mu-\ii\wt\nu,\fr12\pm\ii h\wt\mu+\ii\wt\nu\\
2+p_1\pm\ii h\wt\mu,2+p_2\pm\ii h\wt\mu\edm\right]\n\\
&\pm\ii\Gamma\Big[3+p_{12}\Big] 
 {}_3\wt {\mathrm F}_2\left[\bgm 3+p_{12},\fr12\mp\ii h\wt\mu\pm\ii\wt\nu,1\\ \fr52+p_1\pm\ii\wt\nu,\fr52+p_2\pm\ii\wt\nu \edm\middle| 1 \right]\bigg\}.
\end{align}
\end{keyeqn}
 
\section{Conclusions and Outlooks}
\label{sec_concl}
 
Inflation correlators play central roles in the study of Cosmological Collider physics, and it is also of central importance for a better understanding of quantum field theories in dS. Exact and analytical results for inflation correlators are useful for understanding the analytic structure of dS correlation functions in general, for phenomenological studies of CC physics, and also for efficient numerical implementation. 
 
In this work, we have found exact and closed-form formulae for a wide range of two-point and three-point correlation functions of massless modes in inflationary spacetime mediated by a single massive field at the tree level. As we have shown, tree-level correlators can be easily reduced to (combinations of) seed integrals, either Hankel-type or Whittaker-type. In particular, three/two-point functions correspond to the single/double-folded limit of seed integrals. We calculated the seed integrals using an improved bootstrap method. First, our start point of deriving the bootstrap equation is the equations of motion for the massive propagator, rather than the bulk symmetry, so our method does not depend on full dS symmetry and also applies to dS-boost-breaking models. Second, we found it more convenient to express the bootstrap equations with variables $u_{1,2}$, since the inhomogeneous solution in the single-folded limit can then be summed to a generalized hypergeometric function. With the coefficients of homogeneous solutions appropriately determined, closed-form expressions of seed integrals in the single-folded limit are derived. It is then straightforward to go to the double-folded limit, where all the spurious divergences should be canceled due to the Bunch-Davis initial condition.

Apart from building phenomenological models and constructing efficient templates for practical data analysis, our results also find wide applications for theoretical studies of inflation correlators.
On one hand, low-point tree-level correlators can be subgraphs of more complicated processes,  and our expressions can be used as building blocks. This topic will be explored  in detail in \cite{qin_box}, where the two-point correlators found in this work act as effective vertices in a loop diagram.
On the other hand, closed-form expressions explicitly show the analytical structure of the correlators and thus make the analytic continuation easier to do. The analytic structure of inflation correlators encodes rich physical information, and can provides us useful insights into the physical processes happening in the inflationary universe. They may also pave the way for efficient computation methods for more complicated correlation functions. We leave these topics for future studies.

\paragraph{Acknowledgments.} This work is supported by the National Key R\&D Program of China (2021YFC2203100), NSFC under Grant No.\ 12275146, an Open Research Fund of the Key Laboratory of Particle Astrophysics and Cosmology, Ministry of Education of China, and a Tsinghua University Initiative Scientific Research Program. 
 
\newpage 
\begin{appendix}

\section*{Appendix}
  
\section{Useful Formulae}
\label{app_formulae}
In this appendix, we list some of the definitions and formulae frequently used in the main text.
First, we
use the following shorthand notations for the products and fractions of the Euler $\Gamma$ function:
\begin{align}
\label{eq_GammaProd}
  \Gamma\left[ z_1,\cdots,z_m \right]
  \equiv&~ \Gamma(z_1)\cdots \Gamma(z_m) ,\\
\label{eq_GammaProd2}
  \Gamma\left[\bgm z_1,\cdots,z_m \\w_1,\cdots, w_n\edm\right]
  \equiv&~\FR{\Gamma(z_1)\cdots \Gamma(z_m)}{\Gamma(w_1)\cdots \Gamma(w_n)}.
\end{align}
We also use the Pochhammer symbol $(a)_n\equiv \Gamma(a+n)/\Gamma(a)$ in some expressions. 

The closed-form expressions of two and three-point correlators often involve various types of (generalized) hypergeometric functions. The original (generalized) hypergeometric function is defined as:
\begin{align}
\label{eq_HGF}
  {}_p\mathrm{F}_q\left[\bgm a_1,\cdots,a_p \\ b_1,\cdots,b_q \edm  \middle| z \right]=\sum_{n=0}^\infty\FR{(a_1)_n\cdots (a_p)_n}{(b_1)_n\cdots (b_q)_n}\FR{z^n}{n!}.
\end{align}
We only encounter the case of $p=q+1$, where the above series converges within the disk $|z| < 1$.
At $z=1$, \eqref{eq_HGF} converges if and only if $\Re s > 0$, where the balance $s$ is defined by:
\bge
s = (b_1 + \cdots + b_q) - (a_1 + \cdots a_p).
\ede

When considering the folded limit of the seed integrals, we shall usually take $z\to 1$ for several (generalized) hypergeometric functions. Typically, each of these functions will diverge, but their divergent parts must be canceled with each other. After the cancellation, only the ``finite part" of each (generalized) hypergeometric function remains. Below we list the finite part of the (generalized) hypergeometric functions involved as the argument $z\to 1$ \cite{doi:10.1137/0518089}:
\begin{align}
&\text{Fin}\bigg\{\lim_{z\to 1}{}_2\mathrm F_1 \left[\bgm a,b\\c\edm \middle| z\right]\bigg\} =
\Gamma\left[\bgm c,s\\ a+s,b+s\edm\right],\\
&\text{Fin}\bigg\{\lim_{z\to 1}{}_3\mathrm F_2 \left[\bgm a,b,c\\d,e\edm \middle| z\right]\bigg\} =
\Gamma\left[\bgm d,e,s\\ c,a+s,b+s\edm\right]
{}_3\mathrm F_2 \left[\bgm d-c,e-c,s\\a+s,b+s\edm \middle| 1\right],
\end{align}
both of which hold when the balance $s \notin\mathbb Z$ because of the factor $\Gamma(s)$. When $s$ is an integer, similar formulae can be derived by an approximate limiting process. After some simplifications the expressions in the folded limit will apply for any value of $s$.

We shall also use the regularized and dressed (generalized) hypergeometric functions to simplify expressions, which are respectively defined as the following:
\begin{equation}
\label{eq_RegF}
    {}_p\wt{\mathrm{F}}_q\left[\begin{matrix}
        a_1, \cdots, a_p \\
        b_1, \cdots, b_q
    \end{matrix}\middle|z\right]=\frac1{\Gamma\left[b_1, \cdots, b_q\right]}{}_p\mathrm{F}_q\left[\begin{matrix}
        a_1, \cdots, a_p \\
        b_1, \cdots, b_q
    \end{matrix}\middle|z\right].
\end{equation}
\begin{equation}
\label{eq_DressedF}
    {}_p\mathcal{F}_q\left[\begin{matrix}
        a_1, \cdots, a_p \\
        b_1, \cdots, b_q
    \end{matrix}\middle|z\right]=\Gamma\left[\begin{matrix}
        a_1, \cdots, a_p \\
        b_1, \cdots, b_q
    \end{matrix}\right]{}_p\mathrm{F}_q\left[\begin{matrix}
        a_1, \cdots, a_p \\
        b_1, \cdots, b_q
    \end{matrix}\middle|z\right].
\end{equation}
%
%

\section{Squeezed limit results}
\label{app_squeezed}
In this appendix, we calculate the seed integrals of both Hankel-type \eqref{eq_SeedIntH} and Whittaker-type \eqref{eq_SeedIntW} in a particular squeezed limit, namely $u_1\ll u_2\ll 1$.  The results will help us determine the coefficients of each possible combination of homogeneous solutions to the bootstrap equations.

The full results for seed integrals \eqref{eq_SeedIntH} and \eqref{eq_SeedIntW} have been calculated in \cite{Qin:2022fbv}, using the method of partial Mellin-Barnes representation. We can directly take the squeezed limits of these known expressions to obtain the leading order results. For concreteness, below we repeat a similar calculation, but in a rather brief way, and compute the squeezed limit results. Readers can refer to \cite{Qin:2022fbv} for more details. For convenience, we first calculate the seed integrals defined as \eqref{eq_seedHr} and \eqref{eq_seedIntWr} which depend on variables $r_{1,2}$, and then change to variables $u_{1,2}$. The hierarchical squeezed limit $u_1\ll u_2\ll 1$ is equivalent to $r_1 \ll r_2 \ll 1$. Also notice that we can identify $u_i\simeq 2r_i$ $(i=1,2)$ in the squeezed limit up to $\order{r_{1,2}}$ corrections.

\subsection{Hankel seed integral in squeezed limit}
\label{app_hankel}
The Mellin-Barnes representation for the ``less"/``greater" Wightman function are the following \cite{Qin:2022fbv}:
\begin{align}
D_\lessgtr (k;\tau_1,\tau_2) =&~ \FR{1}{4\pi} \int_{-\ii\infty}^{\ii\infty} \FR{\di s_1}{2\pi\ii} \FR{\di s_2}{2\pi\ii}\, e^{\mp\ii\pi(s_1-s_2)}\Big(\FR k2\Big)^{-2s_{12}}(-\tau_1)^{-2s_1+3/2}(-\tau_2)^{-2s_2+3/2}\n\\
&\times \Gamma\Big[s_1-\FR{\ii\wt\nu}2,s_1+\FR{\ii\wt\nu}2,s_2-\FR{\ii\wt\nu}2,s_2+\FR{\ii\wt\nu}2\Big].
\end{align}
Here and below we use shorthand $s_{12}=s_1+s_2$. The four SK propagators $D_{\aa\bb}$ are related by \eqref{eq_DScalarSame} and \eqref{eq_DScalarOpp}.
Furthermore, we split the same-sign propagators into two parts:
\bge
\label{eq_cuttingrule}
D_{\pm\pm}(k;\tau_1,\tau_2)
  =D_\gtrless(k;\tau_1,\tau_2)+\Big[D_\lessgtr(k;\tau_1,\tau_2)-D_\gtrless(k;\tau_1,\tau_2)\Big]\theta(\tau_2-\tau_1)
\ede
in the region $r_1<r_2$ (equivalently, $u_1<u_2$, and $k_{12}>k_{34}$).
The two parts will give rise to the factorized (F) part ${\mathcal I}_{\pm\pm,\text{F},>}^{p_1p_2}$ and time-ordered (TO) part ${\mathcal I}_{\pm\pm,\text{TO},>}^{p_1p_2}$ of the same-sign seed integrals $\mathcal I_{\pm\pm}(r_1,r_2)$, respectively. The subscript $>$ serves as a reminder that we take $r_1<r_2$ and hence $k_{12}>k_{34}$. Naively, we can split $D_{\pm\pm}$ in a different way, with the factorized part being $D_\lessgtr$. This choice will lead to an exchange of $r_1 \leftrightarrow r_2$ in the expressions \eqref{eq_ScaSqueezedOpp}, \eqref{eq_ScaSqueezedSameF} and \eqref{eq_ScaSqueezedSameTO}, and the time-ordered part will be non-analytic in the limit $r_1 \ll r_2$ and contribute even in the leading order. In fact, our choice \eqref{eq_cuttingrule} is the appropriate ``cutting rule" in the region $r_1<r_2$, and it can be easily proved by the Mellin-Barnes representation. See \cite{Qin:2022fbv} for more details, and also \cite{Tong:2021wai} for an intuitive explanation in the bulk. 

Now we insert the Mellin-Barnes representation for the four SK propagators into the definition of seed integral \eqref{eq_seedHr}, complete the trivialized time integrals, and then obtain:
\begin{align}
\label{eq_ScaSqueezedOpp}
 \mathcal{I}_{\pm\mp}^{p_1p_2} =&~ \FR{1}{4\pi}e^{\mp\ii\bar p_{12}\pi/2}r_1^{5/2+p_1}r_2^{5/2+p_2}\int_{-\ii\infty}^{\ii\infty}\FR{\di s_1}{2\pi\ii}\FR{\di s_2}{2\pi\ii}\,
    \Big(\FR{r_1}2\Big)^{-2s_1}\Big(\FR{r_2}2\Big)^{-2s_2}\n\\
    &\times \Gamma\Big[\fr52+p_1-2s_1,\fr52+p_2-2s_2,s_1-\fr{\ii\wt\nu}2,s_1+\fr{\ii\wt\nu}2,s_2-\fr{\ii\wt\nu}2,s_2+\fr{\ii\wt\nu}2\Big],\\
\label{eq_ScaSqueezedSameF}
{\mathcal I}_{\pm\pm,\text{F},>}^{p_1p_2}=&~\FR{1}{4\pi}e^{\mp\ii p_{12}\pi/2}r_1^{5/2+p_1}r_2^{5/2+p_2} \int_{-\ii\infty}^{\ii\infty}\FR{\di s_1}{2\pi\ii}\FR{\di s_2}{2\pi\ii}\,(\pm \ii e^{\pm 2\ii\pi s_1})
    \Big(\FR{r_1}2\Big)^{-2s_1}\Big(\FR{r_2}2\Big)^{-2s_2}\n\\
   &\times \Gamma\Big[\fr52+p_1-2s_1,\fr52+p_2-2s_2,s_1-\fr{\ii\wt\nu}2,s_1+\fr{\ii\wt\nu}2,s_2-\fr{\ii\wt\nu}2,s_2+\fr{\ii\wt\nu}2\Big],\\
   \label{eq_ScaSqueezedSameTO}
    {\mathcal I}_{\pm\pm,\text{TO},>}^{p_1p_2} 
    =&~ \FR{1}{4\pi }e^{\mp\ii\pi(p_1+p_2)/2}r_1^{5+p_{12}}\int_{-\ii\infty}^{\ii\infty}\FR{\di s_1}{2\pi\ii}\FR{\di s_2}{2\pi\ii}\,(\mp \ii e^{\mp 2\ii\pi s_1} \pm \ii e^{\pm 2\ii\pi s_2})
    \Big(\FR{r_1}2\Big)^{-2s_{12}}\n\\
    &\times \Gamma\Big[\fr52+p_2-2s_2,5+p_{12}-2s_{12},s_1-\fr{\ii\wt\nu}2,s_1+\fr{\ii\wt\nu}2,s_2-\fr{\ii\wt\nu}2,s_2+\fr{\ii\wt\nu}2\Big]\n\\
    &\times {}_2\wt{\mathrm{F}}_1\left[\bgm \fr52+p_2-2s_2,5+p_{12}-2s_{12}\\\fr72+p_2-2s_2\edm\middle|\,-\FR{r_1}{r_2}\right].
\end{align}
The last step is to complete the above Barnes-type integrals using the residue theorem. Since we focus on the region $r_1<r_2<1$, we should close the contour from left, and sum over residues at the left poles for both $s_1$ and $s_2$:
\bge
s_i = -n_i \mp \ii\FR{\wt\nu}2, \qquad n_i \in \mathbb N,\quad i=1,2,
\ede
then we can obtain the full result.

However, we only need the leading order result in the squeezed limit $r_1 \ll r_2 \ll 1$, which obviously corresponds to the poles with $n_{1,2}=0$. Furthermore, we find both $\mathcal I_{\pm\mp}^{p_1p_2}$ and $\mathcal I_{\pm\pm,\text{F},>}^{p_1p_2}$ are of order $\mathcal O(r_1^{5/2+p_1}r_2^{5/2+p_2})$, but the time-ordered integral $\mathcal I_{\pm\pm,\text{TO},>}^{p_1p_2}$ is of order $\mathcal O(r_1^{5+p_{12}})$ and thus can be neglected since we are working in the region $r_1 \ll r_2$ and we assume $\text{Re}\,p_{2}>-5/2$ as explained below (\ref{eq_SeedIntH}). So finally we can obtain the leading order results by summing the residues at four poles $s_i = \mp \ii \wt\nu/2$:
\begin{align}
\lim_{r_1\ll r_2\ll 1} \mathcal I_{\pm\mp}^{p_1p_2}(r_1,r_2) =&~
\FR{e^{\mp\ii \bar p_{12}\pi/2}}{4\pi} \Big[\wh{\mathcal Y}_+^{p_1}(r_1)+\wh{\mathcal Y}_-^{p_1}(r_1)\Big]\Big[\wh{\mathcal Y}_+^{p_2}(r_2)+\wh{\mathcal Y}_-^{p_2}(r_2)\Big],\\
\lim_{r_1\ll r_2\ll 1} \mathcal I_{\pm\pm}^{p_1p_2}(r_1,r_2) =&~\FR{\pm\ii e^{\mp\ii p_{12}\pi/2}}{4\pi}\Big[e^{\pi\wt\nu}\wh{\mathcal Y}_\pm^{p_1}(r_1)+e^{-\pi\wt\nu}\wh{\mathcal Y}_{\mp}^{p_1}(r_1)\Big]\Big[\wh{\mathcal Y}_+^{p_2}(r_2)+\wh{\mathcal Y}_-^{p_2}(r_2)\Big],
\end{align}
where
\bge
\wh{\mathcal Y}_\pm^{p}(r) = 2^{\mp\ii\wt\nu} r^{5/2+p\pm\ii\wt\nu}\Gamma\Big[\FR52+p\pm\ii\wt\nu,\mp\ii\wt\nu\Big].
\ede
One can straightforwardly write down the above results in terms of variables $u_{1,2}$:
\begin{align}
\lim_{u_1\ll u_2\ll 1}\wt{\mathcal I}^{p_1p_2}_{\pm\mp}(u_1,u_2) 
=&~\FR{e^{\mp\ii\bar{p}_{12}\pi/2}}{4\pi}\Big[\wt{\mathcal Y}_+^{p_1}(u_1)+\wt{\mathcal Y}_-^{p_1}(u_1)\Big]\Big[\wt{\mathcal Y}_+^{p_2}(u_2)+\wt{\mathcal Y}_-^{p_2}(u_2)\Big],\\
\lim_{u_1\ll u_2\ll 1}\wt{\mathcal I}^{p_1p_2}_{\pm\pm}(u_1,u_2) 
=&~\FR{\pm\ii e^{\mp\ii p_{12}\pi/2}}{4\pi}\Big[e^{\pi\wt\nu}\wt{\mathcal Y}_\pm^{p_1}(u_1)+e^{-\pi\wt\nu}\wt{\mathcal Y}_{\mp}^{p_1}(u_1)\Big]\Big[\wt{\mathcal Y}_+^{p_2}(u_2)+\wt{\mathcal Y}_-^{p_2}(u_2)\Big].
\end{align}
where
\begin{align}
 \wt{\mathcal Y}_\pm^p(u)= 2^{\mp \ii\wt\nu} \Big(\FR{u}{2}\Big)^{5/2+p\pm\ii\wt\nu}\Gamma\Big[\FR52+p\pm\ii\wt\nu,\mp\ii\wt\nu\Big].
\end{align}

\subsection{Whittaker seed integral in squeezed limit}
\label{app_whittaker}
 The calculation for the Whittaker seed integral is basically the same as the previous case, except that there are different choices of Mellin-Barnes representation for the propagators. We find that the simplest choice is the following \cite{Qin:2022fbv}:
For the opposite-sign propagators $D_{\pm\mp}^{(h)p_1p_2} = D_{\lessgtr}^{(h)p_1p_2}$, we use:
\begin{align}
D_{\lessgtr}^{(h)}(k;\tau_1,\tau_2)=&~
\FR{e^{-h\pi\wt\mu}}{2\pi^2}(\cosh2\pi\wt\mu+\cosh2\pi\wt\nu)e^{\pm\ii k(\tau_1-\tau_2)}\n\\
&\times\int_{-\ii\infty}^{\ii\infty} \FR{\di s_1}{2\pi\ii}\FR{\di s_2}{2\pi\ii}\,
e^{\mp\ii\pi(s_1-s_2)/2}(2k_s)^{-s_{12}}(-\tau_1)^{-s_1+1/2}(-\tau_2)^{-s_2+1/2}\n\\
&\times \Gamma\Big[-s_1+\fr12\pm \ii h\wt\mu,-s_2+\fr12\mp \ii h\wt\mu,s_1-\ii\wt\nu,s_1+\ii\wt\nu,s_2-\ii\wt\nu,s_2+\ii\wt\nu\Big],
\end{align}
and for the same-sign propagators $D_{\pm\pm}^{(h)p_1p_2}$, we again do the split:
\bge
D_{\pm\pm}^{(h)}(k;\tau_1,\tau_2)
  =D_\gtrless(k;\tau_1,\tau_2)+\Big[D_\lessgtr(k;\tau_1,\tau_2)-D_\gtrless(k;\tau_1,\tau_2)\Big]\theta(\tau_2-\tau_1),
\ede
 and use another representation:
 \begin{align}
D_{\lessgtr}^{(h)}(k;\tau_1,\tau_2)=&~
\FR{e^{-h\pi\wt\mu}}{\pi\Gamma[\fr12-\ii\wt\nu\mp\ii h\wt\mu,\fr12+\ii\wt\nu\mp\ii h\wt\mu]}e^{\mp\ii k(\tau_1+\tau_2)}\n\\
&\times\int_{-\ii\infty}^{\ii\infty} \FR{\di s_1}{2\pi\ii}\FR{\di s_2}{2\pi\ii}\,
e^{\mp\ii\pi(s_1-s_2)/2}\cos\pi(s_1\pm\ii h\wt\mu)(2k_s)^{-s_{12}}(-\tau_1)^{-s_1+1/2}(-\tau_2)^{-s_2+1/2}\n\\
&\times \Gamma\Big[-s_1+\FR12\mp \ii h\wt\mu,-s_2+\FR12\mp \ii h\wt\mu,s_1-\ii\wt\nu,s_1+\ii\wt\nu,s_2-\ii\wt\nu,s_2+\ii\wt\nu\Big].
\end{align}
Similar to the Hankel case, we insert the above Mellin-Barnes representation into the Whittaker seed integral \eqref{eq_seedIntWr}. After integrating out $\tau_{1,2}$, we make use of the residue theorem to compute the Barnes-type integral over $s_{1,2}$, with the left poles:
\bge
s_i = -n_i \mp \ii\wt\nu,\qquad n_i\in\mathbb N, \quad i=1,2.
\ede
Again, the leading order result in the squeezed limit comes from the residues at $s_i = \mp \ii\wt\nu$, and the time-ordered integral is negligible since $r_1\ll r_2$. The results can be summarized collectively in the following form:
\bge
\lim_{r_1\ll r_2\ll 1} \mathcal I_{\aa\bb}^{(h)p_1p_2}(r_1,r_2) = \sum_{\cc,\dd = \pm} \be_{\aa\bb|\cc\dd}\, \wh {\mathcal U}_{\aa|\cc}^{p_1}(r_1)\wh {\mathcal U}_{\bb|\dd}^{p_2}(r_2), 
\ede
where $\be_{\aa\bb|\cc\dd}$ are defined in \eqref{eq_betaCoef1}, \eqref{eq_betaCoef2}, \eqref{eq_betaCoef3}, and
\bge
\wh {\mathcal U}_{\aa|\bb}^p(r) = \ii\,\aa\bb\, 2^{\ii\aa\bb\wt\nu} \pi \text{csch}(2\pi\wt\nu)  \,r^{3/2+p+ \ii\aa\bb\wt\nu}\Gamma\Big[\bgm \fr32+p+\ii\aa\bb\wt\nu,\fr12+\ii\aa h\wt\mu+\ii\aa\bb\wt\nu\\1+2\ii\aa\bb\wt\nu \edm\Big].
\ede
One can then turn to the expression in $u$-variables:
\bge
\label{eq_IhSqLimit}
\lim_{u_1\ll u_2\ll 1} \mathcal I_{\aa\bb}^{(h)p_1p_2}(u_1,u_2) = \sum_{\cc,\dd = \pm} \be_{\aa\bb|\cc\dd}\, \wt {\mathcal U}_{\aa|\cc}^{p_1}(u_1)\wt {\mathcal U}_{\bb|\dd}^{p_2}(u_2), 
\ede
where
\bge
\wt {\mathcal U}_{\aa|\bb}^p(u) = \ii\,\aa\bb\, 2^{\ii\aa\bb\wt\nu} \pi \text{csch}(2\pi\wt\nu)  \Big(\FR u2\Big)^{3/2+p+ \ii\aa\bb\wt\nu}\Gamma\Big[\bgm \fr32+p+\ii\aa\bb\wt\nu,\fr12+\ii\aa h\wt\mu+\ii\aa\bb\wt\nu\\1+2\ii\aa\bb\wt\nu \edm\Big].
\ede
\end{appendix}

\newpage
\bibliography{CosmoCollider} 
\bibliographystyle{utphys}

\end{document}